\documentclass[useAMS,usenatbib,usegraphicx,letterpaper]{mn2e}

\usepackage{multirow} 
\usepackage{epsfig}
\usepackage{subfigure}

\newcommand{\etal}{et al.}
\newcommand{\leqsim}{\raisebox{-0.6ex}{$\,\stackrel
        {\raisebox{-.2ex}{$\textstyle <$}}{\sim}\,$}}

\def\mnras{MNRAS}

\def\apj{ApJ}
\def\araa{Annual Review of Astronomy \& Astrophysics}

\def\apjl{ApJ}
\def\aap{A\&A}
\def\aj{AJ}
\def\pasj{PASJ}
\def\procspie{Proc. SPIE}

\title[Clumpy torus and BLR in ESO~323--G77] {The properties of the
  clumpy torus and BLR in the polar--scattered Seyfert~1 galaxy
  ESO~323--G77 through X--ray absorption variability}

\author[G.\ Miniutti \etal]
       {\parbox{\textwidth}{G.~Miniutti,$^{1}$\thanks{E-mail:
             \texttt{gminiutti@cab.inta-csic.es}} M.~Sanfrutos,$^{1}$
           T.~Beuchert,$^{2}$ B.~Ag\'is-Gonz\'alez,$^{1}$ A.~L.~Longinotti,$^{3}$
           E.~Piconcelli,$^{4}$
           Y.~Krongold,$^{5}$ M.~Guainazzi,$^{3}$ S.~Bianchi,$^{6}$
           G.~Matt,$^{6}$ and
           E.~Jim\'enez-Bail\'on,$^{5}$}\vspace{0.5cm}\\
\parbox{\textwidth}{
$^{1}$Centro de Astrobiolog\'{i}a (CSIC--INTA), Dep. de Astrof\'{i}sica; 
ESAC, PO Box 78, Villanueva de la Ca\~nada, E-28691 Madrid, Spain
\\ 
$^{2}$Dr. Remeis Sternwarte \& ECAP, Universit\"at Erlangen-N\"urberg, Sternwartstr. 7, 96049 Bamberg, Germany
\\
$^{3}$European Space Astronomy Centre of ESA P.O. Box 78, Villanueva de la Ca\~nada, E-28691 Madrid, Spain 
\\ 
$^{4}$Osservatorio Astronomico di Roma (INAF), Via Frascati 33. I-00040 Monteporzio Catone (Roma) Italy
\\
$^{5}$Instituto de Astronom\'ia, Universidad Nacional Aut\'onoma de M\'exico, Apdo. 70-264, Cd. Universitaria, M\'exico DF 04510, M\'exico
\\
$^{6}$Dipartimento di Matematica e Fisica, Universit\`a degli Studi Roma Tre, Via della Vasca Navale 84, 00146 Roma, Italy }}

\pagerange{\pageref{firstpage}--\pageref{lastpage}}
\pubyear{2013}

\usepackage{times}

\begin{document}

\label{firstpage}

\maketitle

\begin{abstract}
We report results from multi--epoch (2006--2013) X--ray observations
of the polar--scattered Seyfert~1 galaxy ESO~323--G77. The source
exhibits remarkable spectral variability from months to years
timescales. The observed spectral variability is entirely due to
variations of the column density of a neutral absorber towards the
intrinsic nuclear continuum. The column density is generally
Compton--thin ranging from a few times $10^{22}$~cm$^{-2}$ to a few
times $10^{23}$~cm$^{-2}$. However, one observation reveals a
Compton--thick state with column density of the order of $1.5\times
10^{24}$~cm$^{-2}$. The observed variability offers a rare opportunity
to study the properties of the X--ray absorber(s) in an active
galaxy. We identify variable X--ray absorption from two different
components, namely (i) a clumpy torus whose individual clumps have a
density of $\leq 1.7\times 10^8$~cm$^{-3}$ and an average column
density of $\sim 4\times 10^{22}$~cm$^{-2}$, and (ii) the broad line
region (BLR), comprising individual clouds with density of
$0.1-8\times 10^9$~cm$^{-3}$ and column density of
$10^{23}-10^{24}$~cm$^{-2}$. The derived properties of the clumpy
torus can also be used to estimate the torus half--opening angle,
which is of the order of $47^\circ$. We also confirm the previously
reported detection of two highly ionized warm absorbers with outflow
velocities of $1000-4000$~km~s$^{-1}$. The observed outflow velocities
are consistent with the Keplerian/escape velocity at the BLR. Hence,
the warm absorbers may be tentatively identified with the warm/hot
inter--cloud medium which ensures that the BLR clouds are in pressure
equilibrium with their surroundings. The BLR line--emitting clouds may
well be the cold, dense clumps of this outflow, whose warm/hot phase
is likely more homogeneous, as suggested by the lack of strong
variability of the warm absorber(s) properties during our monitoring.
\end{abstract}


\section{Introduction}

ESO~323--G77 ($z=0.015$) is an optically bright ($m_{\rm V} =
13.2$~mag) Seyfert~1 galaxy originally reported as an active galactic
nucleus (AGN) by Fairall (1986). The nuclear spectrum indicates
significant reddening with $A_{\rm V} \simeq 1.04$ (Winkler 1992;
Winkler et al. 1992), a property that is often observed in Seyfert~1
galaxies showing a high level of optical polarisation such as
e.g. Mrk~231 and Fairall~51 (see Smith et al. 1995; Schmid et
al. 2001). Indeed, Schmid et al. (2003) report the detection of high
linear polarisation in ESO~323--G77, ranging from $\sim 2.2$~\% at
8300~\AA\ to $\sim 7.5$~\% at 3600~\AA\ for the continuum. Similar
amounts of polarisation are found for the broad emission lines. The
position angle of the polarisation does not depend on wavelength, and
it is found to be perpendicular to the orientation of the
[O~\textsc{iii}] ionization cone of the galaxy. This suggests that
ESO~323--G77 is observed at an intermediate inclination of $\sim
45^\circ$ with respect to the symmetry axis. Hence, ESO~323--G77
appears to be a borderline Seyfert~1 galaxy, seen at an inclination
that is intermediate between the typical orientation of Seyfert~1 and
Seyfert~2 galaxies, a viewing angle that is likely grazing the
obscuring medium (the so--called torus of Unified models, Antonucci
1993).

If this is the case, then one generally expects that more
centrally--concentrated emitting regions (i.e. X--rays) will show a
higher level of obscuration than more extended ones (e.g. optical and
broad emission lines) as the latter (e.g. their far side) correspond
to lower observer inclination than the former. Indeed, the optical
extinction towards ESO~323--G77, as obtained by Winkler (1992) from
the observed Balmer decrement, corresponds to a column density of
$2.3\times 10^{21}$~cm$^{-2}$ (e.g. G{\"u}ver \& {\"O}zel 2009), much
lower than the X--ray--derived one, which is more than one order of
magnitude higher (Jim\'enez--Bail\'on et al. 2008a), a clear case of
mismatch between optical and X--ray absorption (see Maiolino et
al. 2001). As such, ESO~323--G77 represents an ideal laboratory to
study the properties of the torus and of its atmosphere in the X--ray
band. Here we present results from multi--epoch X--ray observations of
ESO~323--G77 in the 2006--2013 time--frame when the source has been
observed with {\it XMM--Newton}, {\it Swift}, {\it Chandra}, and {\it
  Suzaku}. We focus here on the properties of the broadband X--ray
continuum and on the X--ray absorption measurements. A (on--going)
more detailed analysis of the high--resolution {\it Chandra} and {\it
  XMM--Newton} data devoted to the study of the warm absorber in
ESO~323--G77 will be presented elsewhere (Sanfrutos et al. in
preparation).

\begin{table}
\caption{Reference observation number, X--ray mission, detector,
  observation (starting) date, and net spectral exposure of all X--ray
  observations used in this work. The four Chandra observations have
  been merged, as no significant spectral variability was observed,
  see Section~\ref{obs}.}
\label{tab1}      
\begin{center}
\begin{tabular}{l l l l l}
\hline\hline                 
\# & Mission & Detector & Date & Net exp.~[ks] \\
\\
1 & XMM           & EPIC pn  &2006-02-07 & $\sim$~23 \\
2 & Swift         & XRT~PC    &2006-06-28 & $\sim$~2 \\
3 & Swift         & XRT~PC    &2006-08-17 & $\sim$~4 \\
4 & Swift         & XRT~PC    &2006-09-14 & $\sim$~2 \\
\hline
5 & Chandra  & MEG/HEG   &2010-04-14 & $\sim$~46 \\
6 & Chandra  & MEG/HEG   &2010-04-19 & $\sim$~118 \\
7 & Chandra  & MEG/HEG   &2010-04-21 & $\sim$~60 \\
8 & Chandra  & MEG/HEG   &2010-04-24 & $\sim$~67 \\
\hline
9 & Suzaku        & XIS(0,3)/PIN &2011-07-20 & $\sim$~88 \\
10 & XMM          & EPIC pn  &2013-01-17 & $\sim$~89 \\ 
\hline\hline                        
\end{tabular}
\\
\end{center}
\end{table}

\section{Observations}
\label{obs}

X--ray pointed observations of ESO~323--G77 have been performed with
{\it XMM--Newton} (Jansen et al. 2001), {\it Swift} (Gehrels et
al. 2004), {\it Chandra} (Weisskopf et al. 2000), and {\it Suzaku}
(Mitsuda et al. 2007) since 2006. Details on the X--ray observations
used in this work are given in Table~\ref{tab1}. The data from the
various missions and detectors have been reduced as standard using the
dedicated software {\small SAS v12.01} ({\it XMM--Newton}), {\small
  CIAO v4.4} ({\it Chandra}), and {\small HEASOFT v6.11} ({\it Swift}
and {\it Suzaku}). Epoch-- and position--dependent ancillary responses
and redistribution matrices have been generated for each data
set. Source products have been extracted from circular regions centred
on the source, and background products have been generated from
source--free regions close to the source. For simplicity, we only
consider here EPIC pn spectra from {\it XMM--Newton}, although we have
checked their consistency with the MOS data. The data from all EPIC
cameras are in excellent agreement in the 0.5--10~keV band. On the
other hand, as the MOS and pn data exhibit some discrepancy below
0.5~keV, we make a conservative choice and we use the EPIC pn data
in the 0.5--10~keV band. As for the {\it Suzaku} data, we merge the
spectra from the front--illuminated CCD detectors XIS0 and XIS3 using
the {\small FTOOL ADDASCASPEC} (unless otherwise specified). We also
make use of the PIN data from the Suzaku HXD, which have been reduced
with the dedicated {FTOOL HXDGSOXBPI}. {\it Chandra} observed the
source on 4 separate occasions (see Table~\ref{tab1}) with the
High--Energy--Transmission--Gratings (HETG). As our study focus on the
continuum properties rather than on the superimposed spectral
features, we are in principle interested in the higher
signal--to--noise zeroth--order spectra. We have then extracted
zeroth--order ACIS spectra from all observations. However, we find
that the number of counts per frame in the innermost $3\times 3$~pixel
region exceeds the pile--up limit for 10~\% pile--up in all
observations. Hence, we use here the gratings spectra from the MEG and
HEG detectors that are pile--up free. Moreover, after checking that
there are no significant spectral differences between the individual
{\it Chandra} observations (5, 6, 7, and 8 in Table~\ref{tab1}), we
merge all {\it Chandra} gratings spectra, to obtain one single $\sim
10$~days--averaged spectrum for each grating, representative of epoch
2010-04-14~to~2010-04-24 with a net exposure of $\sim 291$~ks. All CCD
spectra are grouped to a minimum of 25 counts per
background--subtracted energy bin, except the {\it Swift} XRT spectra
that are grouped to a minimum of 5 counts per background--subtracted
energy bin (we use the C--statistic for the spectral analysis of the
{\it Swift} data). The {\it Chandra} MEG/HEG gratings spectra have
been grouped to 30 channels per bin, and we checked that this grouping
enables us to use the $\chi^2$ statistic, as for all other data except
the {\it Swift} XRT ones.

\begin{figure}
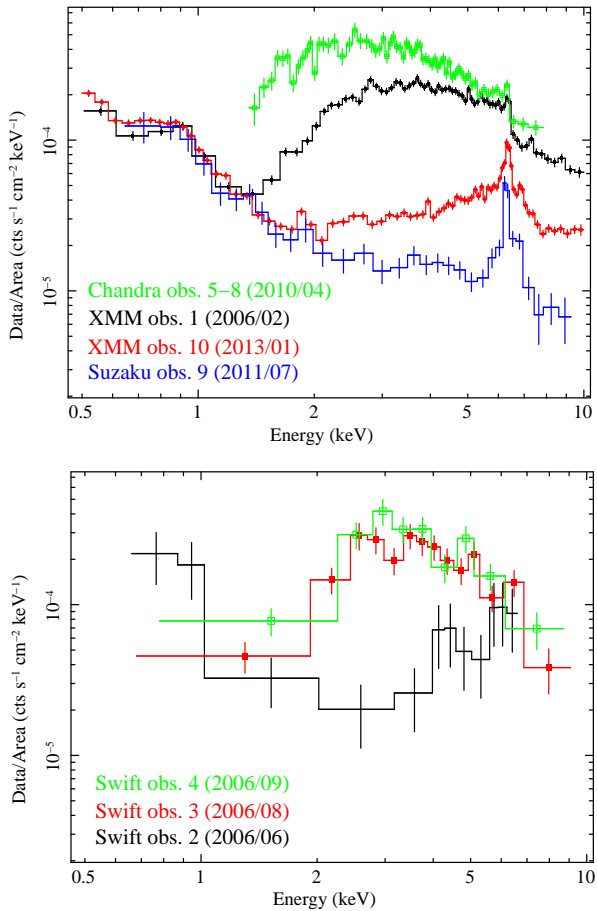

\begin{center}
\includegraphics[width=0.33\textwidth,height=0.45\textwidth,angle=-90]{aaa.ps}
{\vspace{0.3cm}}
\includegraphics[width=0.33\textwidth,height=0.45\textwidth,angle=-90]{SwiftObs.ps}
\caption{In the upper panel, we show the range of X--ray absorption
  variability in ESO~323--G77 as observed with {\it XMM--Newton}, {\it
    Chandra} (only the HEG data are shown for clarity), and {\it
    Suzaku} XIS. In the bottom panel we show the remaining {\it Swift}
  observations. All spectra have been rebinned for visual clarity.}
\label{VarRange}
\end{center}
\end{figure}

\section{Remarkable X--ray absorption variability in ESO~323--G77}

The X--ray spectra of all observations reported in Table~\ref{tab1}
are shown in Fig.~\ref{VarRange}. Each spectrum has been divided by
the corresponding detector effective area to ease visual
comparison. In the upper panel of Fig.~\ref{VarRange}, we show the
full variability range that is observed in ESO~323--G77 from the
higher quality observations in Table~\ref{tab1}. Visual inspection of
the spectra strongly suggests that the observed spectral variability
is mostly associated with X--ray absorption changes. In the bottom
panel of the same Fig.~\ref{VarRange}, we show the remaining {\it
  Swift} observations (2,3, and 4 in Table~\ref{tab1}). The reason to
show them in a different panel is twofold: firstly, the spectral shape
during the {\it Swift} observations is almost coincident with that
during the {\it XMM--Newton} ones (observations 2 and 10 are similar,
as are observations 3, 4, and 1), so that the {\it Swift} data are shown
separately for clarity; secondly, the {\it Swift} observations reveal
the fastest variability event of the overall monitoring and are then
worth special attention: as seen in the bottom panel of
Fig.~\ref{VarRange}, observation 2 (black) is significantly more
absorbed than observation 3 (red) which was performed $\sim 2$~months
after.

\subsection{Looking for a global spectral model}

In order to define a global spectral model to be applied to all
multi--epoch data, we first consider the two {\it XMM--Newton}
observations 1 and 10 (see Table~\ref{tab1}) which have the highest
spectral quality. Hereafter, we consider the X--ray data in the
0.5--10~keV band, unless specified otherwise. Spectral analysis is
performed using the {\small{XSPEC}} v12.7 software (Arnaud 1996). We
start our analysis by defining a phenomenological, typical
Seyfert~2--like spectral model comprising, besides Galactic absorption
(applied to all components with $N_{\rm H}=7.98\times
10^{20}$~cm$^{-2}$, see Kalberla et al. 2005), an intrinsic AGN power
law nuclear continuum absorbed by a column density
$N^{\rm{nucl}}_{\rm{H}}$ at the redshift of ESO~323--G77, a Gaussian
emission line at rest--frame $\sim 6.4$~keV describing Fe K$\alpha$
emission, and a soft power--law which is only absorbed by the Galactic
column density and represents the scattered component that is often
associated with soft X--ray emission lines in the X--ray spectra of
obscured AGN (e.g. Bianchi et al. 2005; Guainazzi \& Bianchi 2007).

\begin{figure}
\begin{center}
\includegraphics[width=0.33\textwidth,height=0.45\textwidth,angle=-90]{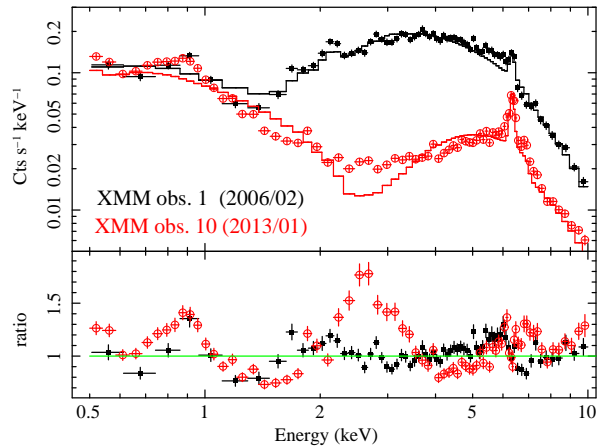}
\caption{Data and models (upper panel), and data--to--model ratio
  (lower panel) for the {\it XMM--Newton} observations 1 and 10 using
  a typical Seyfert~2--like X--ray spectral model comprising an
  absorbed X--ray continuum, a narrow Fe K$\alpha$ emission line, and
  a soft X--ray power law component representing scattering/emission
  from distant material.}
\label{ratiobad}
\end{center}
\end{figure}

We force the continuum spectral index to be the same during both
observations, deferring a discussion about this hypothesis to a later
stage. As soft X--ray emission in obscured AGN likely originates in an
extended region (possibly the narrow line region, see Bianchi,
Guainazzi \& Chiaberge 2006, as well as star--forming regions in the
host galaxy), we force the soft power law normalisation to be the same
at both epochs, and we also assume the same spectral index as that of
the continuum. All the other parameters are allowed to vary
independently. The model, which can be thought of as a rather typical
Seyfert~2 spectral model, is not successful in describing the data
resulting into  $\chi^2 = 2343$ for 1450 degrees of freedom
(dof). The data, spectral models, and data--to--model ratio for both
observations are shown in Fig.~\ref{ratiobad}. Important residuals are
left in both observations around 0.9~keV pointing to the presence of a
further soft (and constant) emission component. On the other hand, the
two spectra differ significantly around 2--4~keV and in the Fe K
band. In particular, the overall hard X--ray spectral shape of the more
absorbed {\it XMM--Newton} observation 10 is not at all well described
by our model.

As a first step, in order to account for the common residuals around
0.9~keV, we add an optically--thin plasma emission model, representing
phenomenologically the soft X--ray contribution from either
photo--ionized or collisionally--ionized gas (we use the
{\small{APEC}} model in {\small{XSPEC}}, Smith et al. 2001). The
normalisation is assumed to be the same at the two epochs, as the
model is physically associated with extended emission. The fitting
statistic improves significantly with $\chi^2= 2154$ for 1448 dof for
a plasma temperature of $kT^{\rm{(1)}} \simeq 0.7$~keV. Adding a
second {\small{APEC}} component results in a better statistic with
$\chi^2= 2019$ for 1446 for a temperature of $kT^{\rm{(2)}} \simeq
0.1$~keV. These components account for the residuals below $\sim
2$~keV in Fig.~\ref{ratiobad} almost completely.

As already pointed out by Jim\'enez--Bail\'on et al. (2008a), the {\it
  XMM--Newton} observation 1 exhibits blueshifted absorption lines in
the Fe K band that these authors model with two highly ionized
outflowing absorbers. We then first add one ionized, outflowing
absorber to our overall model using the {\small{ZXIPCF}} model (Reeves
et al. 2008) in {\small{XSPEC}} assuming a covering fraction of
unity. For simplicity, we force the same absorber parameters (column
density and ionization) at both epochs, deferring the analysis of the
ionized absorber variability to a later stage, once a satisfactory
global spectral model is found. The fit improves significantly with
$\chi^2= 1936$ for 1443. Adding the second highly--ionized absorber
identified by Jim\'enez--Bail\'on et al. (2008a) further improves the
statistic to $\chi^2= 1875$ for 1440. The absorbers are both
highly--ionized ($\log\xi \simeq 3.4$ and $\simeq 4.1$) and outflowing
with velocities of $1000-4000$~km~s$^{-1}$ (measured by letting the
redshift parameter free to vary), in good agreement with previous
results.

\begin{figure}
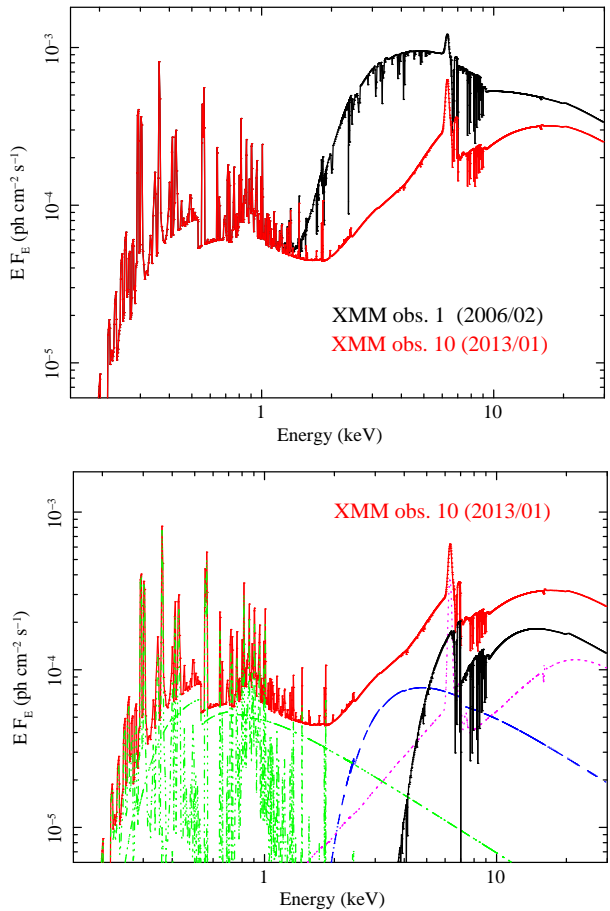

\begin{center}
\includegraphics[width=0.33\textwidth,height=0.45\textwidth,angle=-90]{modelsxmm1and10.ps}
{\vspace{0.3cm}}
\includegraphics[width=0.33\textwidth,height=0.45\textwidth,angle=-90]{modelxmm10.ps}
\caption{The best--fitting models for {\it XMM--Newton} observations 1
  and 10 are shown in the upper panel and are extrapolated to higher
  energies than the data (i.e. to energies $\geq 10$~keV) for
  clarity. In the lower panel, we re--plot the total best--fitting
  model for the more absorbed {\it XMM--Newton} observation 10, but
  include also all spectral components. The soft X--ray band below
  $\sim 2$~keV is described by extended (constant) emission comprising
  a power law contribution and optically--thin emission from
  collisionally--ionized plasma (dot--dashed lines, green in the
  on--line version). The hard X--ray band comprises contributions from
  (i) an heavily absorbed power law representing the intrinsic nuclear
  continuum (solid black line); (ii) an unabsorbed X--ray reflection
  component from optically--thick matter (dotted line, magenta in the
  on--line version); (iii) and additional power law contributing
  mainly in the 2--4~keV band and absorbed by a lower column density
  than the intrinsic nuclear continuum (dashed line, blue in the
  on--line version).}
\label{models}
\end{center}
\end{figure}

As for the emission components in the Fe K band, the Fe K emission is
generally associated with an X--ray reflection continuum. We then
remove the Gaussian line and add a {\small{PEXMON}} model (Nandra et
al. 2007) which self--consistently describes the reflection continuum
and most important emission lines arising from an illuminated slab of
optically--thick gas\footnote{The {\small{PEXMON}} model links the
  reflection continuum to the Fe line intensity assuming a disc--like
  geometry. As shown e.g. by Murphy \& Yaqoob 2009, this is not always
  appropriate for a torus--like geometry. However, using two separate
  models for the Fe emission line and the reflection continuum gives
  exactly the same results as the {\small{PEXMON}} model, with the sum
  of the Fe line and reflection continuum models being coincident with
  the overall {\small{PEXMON}} model we use. In summary, there are no
  strong indications in the data against the use of the
  {\small{PEXMON}} model.}. As our spectral fits suggest that the Fe
emission line has constant intensity with $I_{\rm{Fe,obs.~1}}= (1.6\pm
0.4)\times 10^{-5}$~ph~cm$^{-2}$~s$^{-1}$ and $I_{\rm{Fe,obs.~10}}=
(1.8\pm 0.2)\times 10^{-5}$~ph~cm$^{-2}$~s$^{-1}$, we force all
{\small{PEXMON}} parameters to be the same at both epochs. We convolve
the reflection model with a Gaussian kernel to allow for some
non--zero width of the emission lines. The illuminating spectral index
is set to be the same as that of the intrinsic continuum, we assume
Solar abundances and an inclination of $45^\circ$, so that the only
free parameter of the reflection model is its normalisation (which is
the same at both epochs). As we assume constant reflection, the
reflection fraction $R$ varies with the source intrinsic flux. As an
example, we here report its value ($R\sim 0.65$) for observation 1
only. The statistical description of the data improves to
$\chi^2=1750$ for 1442 dof (to be compared with $\chi^2=1875$ for 1440
dof when the simpler Gaussian emission--line model was used). However,
as for all the models used above, the 2--4~keV residuals of
observation 10 shown in Fig.~\ref{ratiobad} are not accounted for.

\begin{figure}
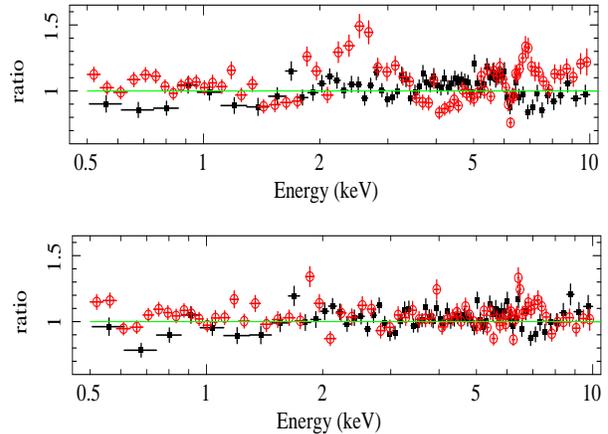

\begin{center}
\includegraphics[width=0.16\textwidth,height=0.45\textwidth,angle=-90]{ratio0.ps}
{\vspace{0.2cm}}
\includegraphics[width=0.16\textwidth,height=0.45\textwidth,angle=-90]{ratio1.ps}
\caption{In the upper panel we show the data--to--model ratio for
  observations 1 and 10 before including the additional absorbed power
  law component ($\chi^2=1750$ for 1442 dof). In the lower panel, we
  show the final, best--fitting data--to--model ratio for both
  observations ($\chi^2=1499$ for 1439 dof).}
\label{ratios}
\end{center}
\end{figure}

After a series of further tests\footnote{Replacing the neutral
  {\small{PEXMON}} reflection model with a ionized one (the
  {\small{REFLION}} model by Ross \& Fabian 2005) and/or allowing for
  a partial covering solution for the neutral absorber does not
  reproduce the data well.}, we find that the data can be described at
both epochs with the addition of a further power law absorbed by a
column density different than that the primary one, whose physical
nature is discussed in Section~\ref{scattexpla}. The normalisation of
this component is allowed to vary between the two epochs. The
statistical description of the data is now excellent with a
$\chi^2=1495$ for 1438 dof. As the additional power law is absorbed by
two column densities that are consistent with each other in the two
observations (and of the order of $7-8 \times 10^{22}$~cm$^{-2}$), we
force the column density to be the same at both epochs which results
into $\chi^2=1499$ for 1439 dof. The best--fitting models for both
observations are shown in the upper panel of Fig.~\ref{models}. In the
lower panel, we show the total best--fitting model for the more
absorbed {\it XMM--Newton} observation 10, together with all spectral
components. In Fig.~\ref{ratios}, we compare the residuals obtained
without (upper panel) and with (lower panel) the inclusion of the
additional absorbed power law component. Some residuals are present
around 6--7~keV, a possible signature of an extra--component or of
some inadequacy of the {\small{ZXIPCF}} absorber model. However, we
refrain from modelling these residuals by adding further ad--hoc
absorption/emission lines to an already complex model. The overall
reduced $\chi^2$ is already $1.04$ so that adding more complexity to
the spectral model would imply the risk of over--modelling the data
(see, however, Section~\ref{broadFe} for a further discussion).

Having found a good description of the X--ray spectrum of
ESO~323--G77, we can now critically consider some of the assumptions
we made. We first allow the intrinsic spectral index to be
different. However, the two spectral indices are consistent with each
other with $\Gamma_{\rm{obs.~1}} = 1.98 \pm 0.08$ and
$\Gamma_{\rm{obs.~10}} = 1.96 \pm 0.07$. We then consider possible
variations in the warm absorber parameters. We keep the outflow
velocities linked between the two epochs, and we first allow for
variations in the ionization parameter only, but no improvement is
obtained. The same holds for variations in the warm absorbers column
density. We must point out, however, that the warm absorbers have
little effect on the more absorbed observation 10, so that their
properties are in fact mostly set by the less absorbed observation
1. Hence, we conclude that the only significant difference between the
two spectra is a one order of magnitude variation in the column
density towards the nuclear continuum (with $N_{\rm H}^{\rm{nucl}}
\simeq 5-6 \times 10^{22}$~cm$^{-2}$ during the {\it XMM--Newton}
observation 1 and $N_{\rm H}^{\rm{nucl}} \simeq 6 \times
10^{23}$~cm$^{-2}$ during observation 10). The best--fitting
parameters will be reported later on, when the multi--epoch spectral
results are presented (Section~\ref{multi--epoch}).

\subsection{Short--timescale variability}
\label{varsec}

\begin{figure}
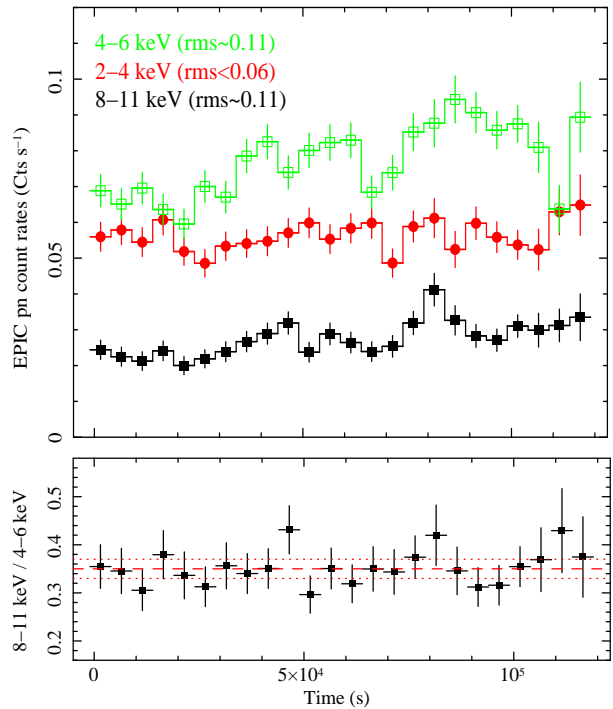

\begin{center}
\includegraphics[width=0.33\textwidth,height=0.45\textwidth,angle=-90]{comp3.ps}
{\vspace{0.0cm}}
\includegraphics[width=0.2\textwidth,height=0.45\textwidth,angle=-90]{rationew3.ps}
\caption{In the upper panel, we show the he {\it XMM--Newton}
  observation 10 EPIC pn light curves in the 4--6~keV (upper curve),
  2--4~keV (middle), and 8--11~keV (lower) energy bands. The bin size
  is 5000~s. In the lower panel of the figure, we show the ratio
  between the two continuum--dominated energy bands of 8--1~keV and
  4--6~keV together with the best--fitting constant and its 90 per
  cent uncertainties ($0.35\pm 0.02$).}
\label{shortvar}
\end{center}
\end{figure}

As clear from the lower panel of Fig.~\ref{models}, our spectral model
predicts no short--timescale variability below $\sim 2$~keV as all
soft components are expected to originate in distant material. On the
other hand, the intrinsic, absorbed, nuclear continuum dominates in
the 4--10~keV band so that some variability may be in principle
expected. As for the additional absorbed power law, this component
dominates in the intermediate 2--4~keV band. The {\it XMM--Newton}
observation 10 is long enough ($\sim 120$~ks) to investigate the
short--timescale variability in different energy bands. We use light
curves with 5000~s bins to compute the rms variability amplitude in
the following energy bands: 0.5--1~keV, 1--2~keV, 2--4~keV, 4--6~keV,
and 8--11~keV. We avoid the 6--8~keV band because this spectral range
is not dominated by one single component, bearing contributions from
continuum, reflection, and from the additional power law. All rms
variability amplitudes are consistent with zero, except that in the
4--6~keV and 8--11~keV bands that are dominated by the absorbed
intrinsic continuum. For those bands, we measure a rms variability
amplitude of $\sim 0.11\pm 0.04$. Therefore, the variability
properties provide support to our spectral model, indicating that
short--timescale variability is only present in energy bands that are
dominated by the nuclear continuum.  Moreover, the energy--dependent
variability also shows that the additional absorbed power law
dominating the 2--4~keV band (see Fig.~\ref{models}, lower panel) is
less variable than the intrinsic continuum. This is shown in the top
panel of Fig.~\ref{shortvar}, where the light curves in the
continuum--dominated 4--6~keV and 8--11~keV energy bands band (upper
and lower curves) are compared with that in the 2--4~keV band (middle
curve), which is dominated by the additional absorbed power law
component.

The observed short--timescale variability in the continuum--dominated
bands can also be used to investigate whether intra--observation
X--ray absorption variability is present. As seen in the lower panel
of Fig.~\ref{models}, the 8--11~keV band is (almost) insensitive to
column density variations, while the softer 4--6~keV band is instead
sensitive to variations of $N_{\rm H}^{\rm{nucl}}$. Hence, the ratio
between the two energy bands can highlight any intra--observation
spectral variability due to absorption changes. The ratio of the
8--11~keV to the 4--6~keV band light curves is shown in the lower
panel of Fig.~\ref{shortvar}. No spectral variability is detected, and
the ratio is consistent with being constant throughout the observation
(a fit with a constant produces $\chi^2=16$ for 23 dof). We then
conclude that no intra--observation absorption variability is detected
during the 120~ks exposure of observation 10.

\subsection{Physical interpretation for the additional absorbed power law component}
\label{scattexpla}

The best--fitting spectral components shown in the lower panel of
Fig.~\ref{models} are all typical of the X--ray spectra of AGN except
the additional absorbed power law component that we introduced to
account for the 2--4~keV residuals during the more absorbed {\it
  XMM--Newton} observation 10 (see e.g. Fig.~\ref{models} and
Fig.~\ref{ratios}). Hence, the inclusion of such additional component
deserves some more detailed physical explanation. We interpret this
component as scattering of the nuclear continuum in a clumpy
absorber. Indeed, if the absorber is clumpy, the observed continuum
comprises a transmitted component, namely an absorbed power law with
column density set by that of the particular clump (or clumps) that
happens to be on our line of sight (LOS), plus a scattering component
due to clumps out of the LOS that intercept and scatter the nuclear
continuum, re--directing part of it into our LOS. As this component
likely arises from a number of different clumps, the scattered
component is absorbed by the average column density of the clumpy
absorber rather than that of one particular clump, which explains why
the column density towards the transmitted and scattered components
may be different and why the column density towards the scattered
component is consistent with being always the same (see e.g. Yaqoob
2012 for a detailed discussion of the envisaged geometry, in
particular his Fig.~2 and Section~7). Modelling the scattered component
with a simple absorbed power law is only a zeroth--order approximation
of the likely spectral shape which encodes our ignorance of the true
scattering geometry. Notice that our scattering interpretation is
consistent with the lower variability amplitude (if any is present) of
the 2--4~keV band with respect to energy bands that are dominated by
the continuum, as scattering in a relatively extended medium tends to
wash out any intrinsic variability.

If our interpretation is correct, the scattering component is likely
to be a fixed fraction of the intrinsic continuum, representing the
scattering efficiency for a given geometry and physical condition of
the scattering medium. To test this further hypothesis we modify our
best--fitting model for the {\it XMM--Newton} observations 1 and 10
forcing the scattered continuum to be a constant fraction of the
intrinsic nuclear continuum in both observations. We indeed find that
the scattered component is consistent with a constant fraction ($14\pm
3$~\%) of the nuclear continuum, with a final statistic of $\chi^2 =
1500$ for 1440 dof, to be compared with $\chi^2 = 1499$ for 1439 dof
when the scattered component is free to vary independently. As the two
fits are statistically equivalent, we retain this further assumption
in subsequent analysis. In summary, our interpretation suggests the
presence of a variable, clumpy absorber whose properties can be investigated in
great detail thanks to the remarkable absorption variability of
ESO~323--G77 (see Fig.~\ref{VarRange}). The remaining part of this
paper is devoted precisely to this goal.

\subsection{The Compton--thick nature of the Suzaku observation}
\label{suzakuspectra}

\begin{figure}
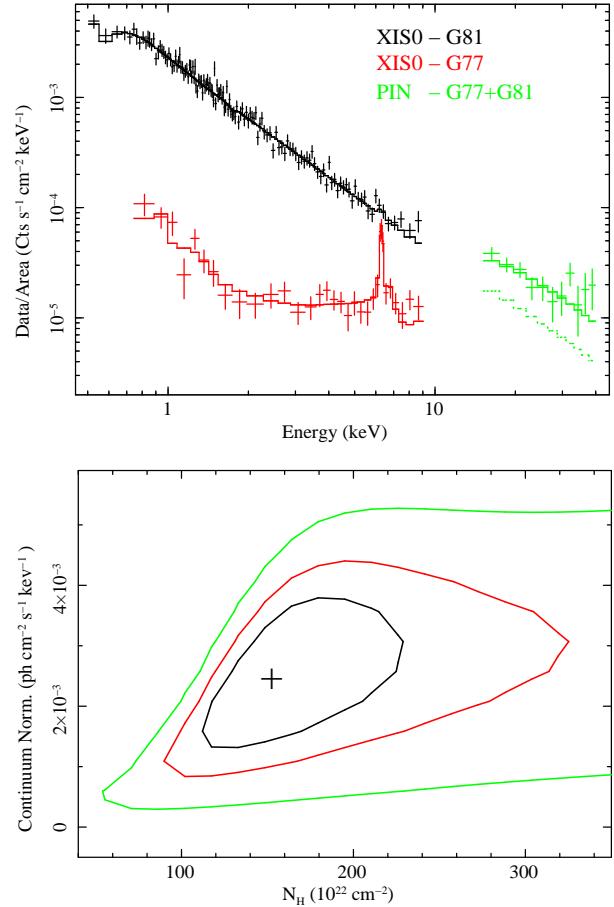

\begin{center}
\includegraphics[width=0.33\textwidth,height=0.45\textwidth,angle=-90]{Suzaku65sec.ps}
{\vspace{0.3cm}}
\includegraphics[width=0.33\textwidth,height=0.45\textwidth,angle=-90]{NHcontnew.ps}
\caption{In the top panel we show the XIS0 spectra of ESO~323--G81 and
  ESO~323--G77 together with their best--fitting models (A and B
  respectively). The PIN spectrum comprises contribution from both
  sources. The best--fitting green solid line is for the A+B model,
  while the dotted line is the extrapolation of the best--fitting
  model B for ESO~323--G81. In the bottom panel, we show the 68, 90,
  and 99 per cent confidence level contours for intrinsic continuum column
  density and normalisation in ESO~323--G77, strongly suggesting a
  Compton--thick state for the {\it Suzaku} observation 9 of
  ESO~323--G77.}
\label{Suzakufigs}
\end{center}
\end{figure}

Before proceeding with a multi--epoch spectral analysis of all
available X--ray observations, we point out that the {\it Suzaku}
observation 9 on 2011/07 deserves some dedicated study. As clear from
Fig.~\ref{VarRange} (upper panel), the Suzaku observation is
reflection--dominated with a very prominent Fe K$\alpha$ emission line
whose equivalent width is $\sim 800$~eV. However, the XIS data (up to
10~keV) cannot discriminate between two different scenarios that can
give rise to reflection--dominated spectra, namely a Compton--thick
state where the nuclear continuum is heavily absorbed only marginally
contributing below 10~keV, or a switched--off state where the nuclear
continuum is extremely dim, only leaving the reprocessed components
below 10~keV.

In principle, {\it Suzaku} is the ideal observatory to accurately
measure the column density towards absorbed AGN thanks to the presence
of the HXD and, in particular, the PIN detector with good sensitivity
up to tens of keV. However, the large ($34\arcmin\times 34\arcmin$)
field--of--view (FOV) of the HXD implies that the PIN X--ray data are
contaminated by the nearby and relatively X--ray bright AGN
ESO~323--G81, which is only $\sim 8.8\arcmin$ away from
ESO~323--G77. No other relatively bright X--ray source is known within
the HXD FOV. As such, we extract the spectra of both ESO~323--G77 and
ESO~323--G81 from the XIS0 detector\footnote{The reason for using only
  the XIS0 data (and not the XIS3 as well) in this analysis is to not
  introduce further cross--calibration uncertainties between the XIS
  and PIN detectors.} using circular regions of same radius (limited
to $65\arcsec$ due to the position of ESO~323--G81, close to the XIS0
detector edge). We then build an overall spectral model for both
ESO~323--G77 (for which we use the same model described in the
previous section) and for ESO~323--G81 which is well described by a
rather typical Seyfert~1 X--ray model comprising Galactic absorption,
a power law continuum, a distant reflection component, and a soft
excess that we model here phenomenologically with a simple
blackbody. Hereafter, the spectral models for ESO~323--G77 and
ESO~323--G81 are called model A and B respectively.

We then fit the XIS~0 spectra of both sources simultaneously with the
PIN spectrum, which comprises contributions from both. The XIS0
spectrum of ESO~323--G77 is described with model A, the XIS0 spectrum
of ESO~323--G81 with model B, and the PIN spectrum with model A+B. We
fix the PIN/XIS0 cross--calibration constant to the value of 1.16 for
ESO~323--G77, which is appropriate for observations performed at the
XIS nominal position. The position of ESO~323--G81 (close to the west
edge of the XIS detectors) falls closer to the HXD nominal
position\footnote{Maeda et al. 2008,
  JX--ISAS--SUZAKU--MEMO--2008--06}. Hence we use a cross--calibration
constant of 1.18 (the value appropriate for the HXD nominal position)
for ESO~323--G81. However, we must point out that our results are
basically insensitive to this conservative choice.

In the top panel of Fig.~\ref{Suzakufigs} we show the XIS0 spectra of
ESO~323--G77 and ESO~323--G81, together with the PIN spectrum and the
corresponding best--fitting model(s). The dotted line represents the
extrapolation of model B (the best--fitting model for ESO~323--G81) in
the PIN band and highlights the hard X--ray excess that we attribute
to ESO~323--G77. The extrapolation of model B to the PIN band suggests
that ESO~323--G81 has a 15--40~keV flux of $(9.0\pm 1.0)\times
10^{-12}$~erg~s$^{-1}$~cm$^{-2}$. On the other hand, the observed PIN
flux is $(1.99\pm 0.23)\times 10^{-11}$~erg~s$^{-1}$~cm$^{-2}$ in the same band,
i.e. more than a factor of 2 higher. In the bottom panel
Fig.~\ref{Suzakufigs} we show the 68, 90, and 99 per cent confidence
level contours for the absorber column density towards the intrinsic nuclear
continuum in ESO~323--G77 and for the continuum normalisation. Although the systematic
errors due to the extrapolation of model B in the PIN band may affect
the uncertainties, our analysis strongly suggests a nearly
Compton-thick or Compton--thick state for the 2011/07 {\it
  Suzaku} observation of ESO~323--G77 with $N_{\rm H}\geq 9\times
10^{23}$~cm$^{-2}$ at the 90 per cent confidence level for any allowed
intrinsic flux level.

\begin{figure}
\begin{center}
\includegraphics[width=0.65\textwidth,height=0.45\textwidth,angle=-90]{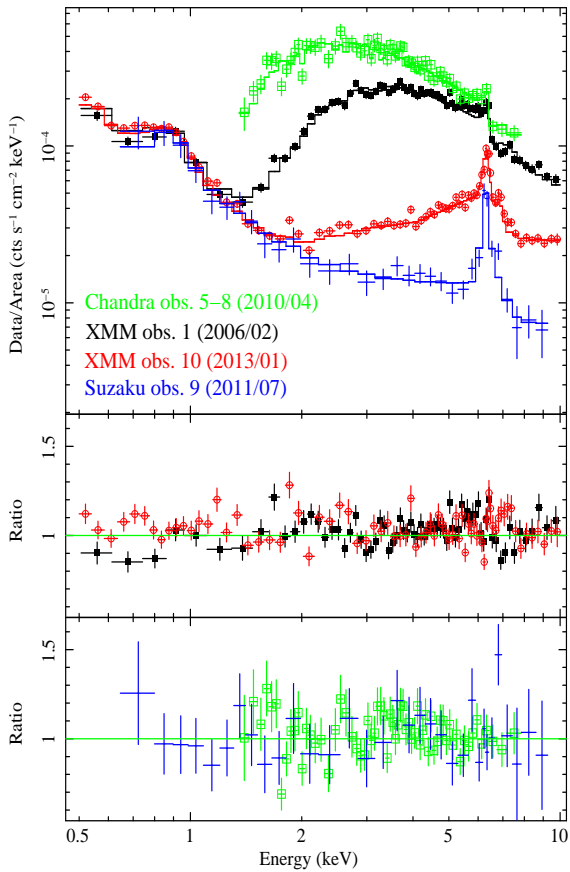}
\caption{Data (divided by the corresponding detector effective area),
  best--fitting model, and data--to--model ratio for the final
  multi--epoch spectral analysis. We only show the highest quality
  observations from {\it XMM--Newton} (observations 1 and 10), {\it
    Chandra} (observation 5--8; only the HEG data are shown), and {\it
    Suzaku} (observation 9) for clarity. The middle panel shows the
  best--fitting data--to--model ratio for the two {\it XMM--Newton}
  observations 1 and 10, while the lower panel is for the {\it
    Chandra} and {\it Suzaku} observations 5--8 and 9.}
\label{res}
\end{center}
\end{figure}

\section{Multi--epoch spectral analysis}
\label{multi--epoch}

\begin{table*}
\caption{Best--fitting parameters from our multi--epoch spectral
  analysis. The final fitting statistic is excellent with
  $\chi^2/{\rm{dof}}=2616/2597$ for the higher quality data from {\it
    XMM--Newton} (obs. 1 and 10), {\it Chandra} (obs. 5--8), and {\it
    Suzaku} (obs. 9), and with $C/{\rm{dof}}= 137/117$ for the lower
  quality {\it Swift} data (obs. 2, 3, and 4).}
\label{tab2}      
\begin{center}
\begin{tabular}{c c c c c c c c}
\hline\hline                 
&\multicolumn{6}{c}{Constant components}&\\
\hline\hline
Continuum &\multicolumn{1}{c}{Soft PL} & \multicolumn{2}{c}{APEC (1)} & \multicolumn{2}{c}{APEC (2)} &  \multicolumn{2}{c}{Reflection}\\
$\Gamma$ & $L_{\rm{0.5-2}}^{\rm{Soft~PL}}$& $kT^{\rm{(1)}}$ & $L_{\rm{0.5-2}}^{\rm{(1)}}$ & $kT^{\rm{(2)}}$ & $L_{\rm{0.5-2}}^{\rm{(2)}}$ & $L_{\rm{2-10}}^{\rm{ref}}$ & $\sigma^{\rm{ref}}$ \\ \\
$1.98 \pm 0.06$& $(6.7\pm 0.3)\times 10^{-2}$ &$0.74\pm 0.05$ &$(2.7\pm 0.2)\times 10^{-2}$ &$0.09\pm 0.02$ &$(9\pm 1)\times 10^{-3}$ &$0.31\pm 0.02$ &$\leq 0.06$  \\
\hline
& \multicolumn{3}{c}{Warm absorber (1)} & \multicolumn{3}{c}{Warm absorber (2)} &   \\
&$N_{\rm H}^{\rm{(1)}}$ & $\log\xi^{\rm{(1)}}$ & $v_{\rm{out}}^{\rm{(1)}}$ &  $N_{\rm H}^{\rm{(2)}}$ & $\log\xi^{\rm{(2)}}$ & $v_{\rm{out}}^{\rm{(2)}}$ & \\ \\
&$9\pm 6$ &$3.4\pm 0.3$ &$3500\pm 700$ &$35\pm 18$ &$4.1\pm 0.3$ &$1500\pm 600$ & \\
\hline\hline                        
&\multicolumn{6}{c}{Variable/continuum components}&\\
\hline\hline
\multicolumn{1}{c}{Obs. \#} & \multicolumn{3}{c}{Intrinsic nuclear continuum} & \multicolumn{2}{c}{Scattered continuum} & \multicolumn{2}{c}{Total}\\
& $N_{\rm H}^{\rm{nucl}}$ & $F_{2-10}^{\rm{nucl}}$ &  $L_{2-10}^{\rm{nucl}}$ & $N_{\rm{H}}^{\rm{scatt}}$ & $L^{\rm{scatt}}/L^{\rm{nucl}}$ & $F_{0.5-2}^{\rm{tot}}$& $F_{2-10}^{\rm{tot}}$ \\ \\
1 &$5.6\pm 0.4$ &$7.3\pm 0.1$ &$6.7\pm 0.4$ &$7.6\pm 0.8$ &$0.15\pm 0.03$ & $0.24\pm 0.01$ & $9.2\pm 0.1$  \\
2 &$50 \pm 18$ &$7\pm 4$ &$5\pm 3$ &'' &'' &$0.19\pm 0.06$ &$3.5\pm 0.8$  \\
3 &$5.4\pm 0.9$ &$7\pm 2$ &$6\pm 2$ &'' &'' &$0.24\pm 0.05$ &$8.3\pm 0.7$ \\
4 &$5\pm 1$ &$8\pm 2$ &$7\pm 2$ &'' &'' &$0.28\pm 0.06$ &$10\pm 1$ \\
5-8 &$2.7\pm 0.3$ &$11.0\pm 0.3$ &$8.4\pm 0.5$ &'' &'' &$0.63\pm 0.01$ &$13.2\pm 0.3$ \\
9 &$150^{+150}_{-60p}$ &$0.4\pm 0.2$ &$3\pm 2$ &'' &'' &$0.16\pm 0.01$ &$0.9\pm 0.1$ \\
10 &$60\pm 7$ &$1.05\pm 0.06$ &$4.5\pm 0.5$ &'' &'' &$0.17\pm 0.01$ &$2.6\pm 0.1$ \\
\hline\hline
\end{tabular}
\\ \raggedright Units: luminosities are unabsorbed and given in units
of $10^{42}$~erg~s$^{-1}$; fluxes are as observed and given in units
of $10^{-12}$~erg~cm$^{-2}$~s$^{-1}$; the temperature of the
{\small{APEC}} components is given in keV and the same units are also
used for the width $\sigma^{\rm{ref}}$ of the Gaussian kernel that is
applied to the reflection model (the upper limit on the kernel width
corresponds to $\leq 6600$~km~s$^{-1}$ in FWHM, consistent with any Fe
line production site from the broad--line--region outwards). Column
densities are given in units of $10^{22}$~cm$^{-2}$, and the warm
absorbers outflow velocities are in units of km~s$^{-1}$. The subscript $p$ for the value of $N_{\rm H}^{\rm{nucl}}$ of observation 9 means that the parameter pegged to its minimum allowed value (see Section~\ref{suzakuspectra}).
\end{center}
\end{table*}

We now apply the spectral model discussed above to the multi--epoch
data, i.e. to all spectra shown in the upper and lower panels of
Fig.~\ref{VarRange}. As the {\it Swift} XRT data only have a minimum
of 5 counts per energy bin, we apply the C--statistic for these data,
and the standard $\chi^2$ minimisation for all the others. We also use
the 90 per cent confidence levels results discussed for the column
density during the {\it Suzaku} observation to impose that
$N_{\rm H}^{\rm{Suzaku}} = (1.5^{+1.7}_{-0.6})\times 10^{24}$~cm$^{-2}$
(see Fig.~\ref{Suzakufigs}, lower panel).

All data are fitted simultaneously in the 0.5--10~keV band (or in
restricted detector--dependent energy bands, depending on the data
quality). In the same way as for the analysis of the two {\it
  XMM--Newton} observations discussed above, most parameters are
allowed to vary during the fit, but are forced to be the same at all
epochs. In particular, we force constant spectral index, soft X--ray
components temperatures and normalisations, reflection flux, and warm
absorbers properties at all epochs. The column density towards the
additional absorbed power law is also forced to be the same at all
epochs, under the assumption that it represents the average clumpy
absorber column density along many different LOS. Moreover, the
(hard) scattered component is assumed to be a constant fraction of the
nuclear continuum, i.e. we force the same scattered fraction at all
epochs. The free parameters that are instead allowed to vary independently are
only the column density towards the intrinsic continuum, and the
continuum normalisation.

We reach an overall excellent description of the data with $\chi^2 =
2616$ for 2597 dof, and $C=137$ for 117 dof. The spectra,
best--fitting models, and data to model ratios for the higher quality
observations from {\it XMM--Newton} (obs. 1 and 10), {\it Chandra}
(obs. 5--8), and {\it Suzaku} (obs. 9) are shown in Fig.~\ref{res},
while the best--fitting parameters are reported in
Table~\ref{tab2}. The most important result of our analysis is the
detection of clear, unambiguous X--ray absorption variability due to
changes of the column density of a neutral absorber towards the
intrinsic nuclear continuum. Table~\ref{tab2} is separated into
constant components (the soft power law and collisionally--ionized
plasmas, the reflection component, and the warm absorbers) and
variable or continuum--related components. We start our discussion
with the former, while the latter will be discussed in
Section~\ref{discuss}.

\section{Results from the spectral analysis}

Multi--epoch X--ray observations of ESO~323--G77 from February 2006 to
January 2013 reveal remarkable spectral variability associated with
X--ray absorption variability. However, the overall spectral model is
complex, and we discuss here the properties of all constant spectral
components, deferring the discussion and interpretation of the
variable absorber to Section~\ref{discuss}.

\subsection{X--ray reflection and soft X--ray emission}

An X--ray reflection component comprising a narrow Fe~K$\alpha$ line is
detected at all epochs. All data are consistent with a constant
reflection (and Fe~K$\alpha$ emission line) flux, so that the observed
Fe~K$\alpha$ line equivalent width varies from $\sim 55$~eV during
the highest flux {\it Chandra} observations 5--8 on 2010/04 to $\sim
800$~eV during the lowest flux {\it Suzaku} observation 9 on
2011/07. According to our spectral model, the X--ray reflection
component is unabsorbed which, together with the constant reflection
flux, suggests reflection from an extended medium with physical size
similar or larger than that of the X--ray absorber(s). We measure a 2--10~keV
reflected luminosity of $L_{2-10}^{\rm{ref}} \sim 3.1\times
10^{41}$~erg~s$^{-1}$, i,.e. about 5~\% of the averaged 2--10~keV
nuclear luminosity of ESO~323--G77 ($\sim 5.8\times
10^{42}$~erg~s$^{-1}$).

The soft X--ray band is here modelled with two constant components,
namely a power--law and collisionally--ionized plasma emission. The
soft X--ray model has to be considered as the simplest
phenomenological description of the data. However, the two constant
components can be identified with (i) scattering of the
nuclear continuum in an extended medium (possibly the narrow line
region, see Bianchi, Guainazzi \& Chiaberge 2006), and (ii) the
contribution of star--forming regions or shocked gas in the AGN
environment and host galaxy. The soft 0.5--2~keV luminosity associated
with the power law is $L_{0.5-2}^{\rm{PL}} \sim 6.7\times
10^{40}$~erg~s$^{-1}$, while the two plasma emission models 
contribute a total luminosity of $L_{0.5-2}^{\rm{apec}} \sim
3.6\times 10^{40}$~erg~s$^{-1}$.

$L_{0.5-2}^{\rm{PL}}$ represents about 1--2~\% of the averaged
intrinsic nuclear luminosity extrapolated to the same band, to be
compared with a 1--5~\% typical soft X--rays scattered fraction in
 obscured AGN (e.g. Bianchi \& Guainazzi 2006). 

On the other hand, the soft X--ray luminosity associated with the
thermal plasma emission components can be used to estimate the
star--formation--rate (SFR) in the galaxy, under the (strong)
assumption that the observed luminosity is due to
star--formation. In fact, as seen in the soft X--ray spectra of many
obscured AGN, part of the emission--line spectrum in the soft X--ray
band is often due to emission from gas that is photo--ionized by the
AGN rather than collisionally--ionized in shocks and/or star--forming
regions (Guainazzi \& Bianchi 2007). Hence, by assuming that the
luminosity of our {\small{APEC}} models is entirely associated with
star--formation, we can only derive an upper limit on the SFR in
ESO~323--G77. By using the relationship ${\rm{SFR}}_{\rm X} \simeq
2.2\times 10^{-40} L_{0.5-2}~M_\odot$~yr$^{-1}$ (Ranalli, Comastri \&
Setti 2003), the X--ray--based estimate of the SFR in ESO~323--G77 is
${\rm{SFR}}_{\rm X} \simeq 6\pm 3 ~M_\odot$~yr$^{-1}$, and thus
${\rm{SFR}}_{\rm X} \leqsim 9 ~M_\odot$~yr$^{-1}$. 

Esquej et al. (2013) derive the SFR in ESO~323--G77 on different
scales, using the $11.3~\rm{µm}$ polycyclic aromatic hydrocarbon (PAH)
feature as a proxy of the SFR.  Using high angular--resolution MIR
spectroscopic observations at the VLT/VISIR (originally presented by
H{\"o}nig et al. 2010), they obtain an upper limit on the nuclear SFR
of ${\rm{SFR}}_{\rm nucl} \leqsim 0.23 ~M_\odot$~yr$^{-1}$ within the
inner $\sim 200$~pc. On the other hand, lower angular--resolution
{\it Spitzer} data were used to estimate an extended SFR of ${\rm{SFR}}_{\rm ext}
\simeq 3 ~M_\odot$~yr$^{-1}$ within $\sim 1$~kpc which is fully
consistent with our ${\rm{SFR}}_{\rm X} \leqsim 9
~M_\odot$~yr$^{-1}$. We point out that the X--ray data are typically
extracted from circular regions of $20-40$~\arcsec radius which
correspond to an encircled radius of $\sim 6-12$~kpc at the distance
of ESO~323--G77.

\subsection{The highly--ionized absorbers}

We confirm the detection of two highly--ionized, outflowing warm
absorbers in ESO~323--G77, as firstly reported by Jim\'enez--Bail\'on
et al. (2008a) in their analysis of the {\it XMM--Newton} observation
1. Although we assume, for simplicity and because the data allow us to
do so, that the parameters of the warm absorbers are the same at all
epochs, it is clear the the absorbers have little, if any, effect on
significantly absorbed observations, and they can only be constrained
reliably from the least absorbed and high quality {\it XMM--Newton}
observation 1 and {\it Chandra} observation 5--8. We then repeat the
analysis by only considering these data sets, in the attempt of
investigating the warm absorbers parameters in some more detail. In
this analysis, we allow the ionization state of the two absorbers to
vary between the different epochs, while keeping outflow velocities
and column densities tied in all data sets. 

The outflow velocities and column densities turn out to be consistent
with those reported in Table~\ref{tab2}. As for the ionization
parameters, only marginal variability is observed, with an overall
$\Delta \chi^2 = 5$ for 2 additional free parameters. On the other
hand, both the observed flux and the intrinsic nuclear luminosity
during the two observations are within a factor of 2 from each other,
and the uncertainties in the warm absorbers ionization parameters do
not allow us to study their response to the continuum variation at
such a fine level (see Table~\ref{tab2}).

Let us consider here the possible origin of such highly--ionized
outflow. One possibility is that the outflow is part of a disc--wind
launched off the inner accretion flow, where radiation pressure is
sufficient to accelerate gas up to escape velocities. As discussed in
Section~\ref{discuss}, typical launching radii are of the order of few
hundreds of $r_g$ (e.g. Risaliti \& Elvis 2010), where the escape
velocity is of the order of $1-3\times 10^4$~km~s$^{-1}$, i.e. about
one order of magnitude higher than the observed outflow velocity of
$1-4\times 10^3$~km~s$^{-1}$. Hence, the wind has to be accelerated
along a direction that is almost perpendicular to our LOS to be
consistent with the relatively low outflow velocity we measure. This
is unlikely, to say the least, given that our LOS has a most probable
inclination of $\sim 45^\circ$ with respect to the symmetry axis
(Schmid et al. 2003; Smith et al. 2004). Assuming a given misalignment
between the outflow direction and our LOS, one can infer limits on the
outflow launching radius $R_{\rm{wind}}$ by assuming that the observed
outflow velocity corresponds to the component along the LOS of the
escape velocity at the launching radius. For $v_{\rm{out}} = 1-4\times
10^3$~km~s$^{-1}$ and a LOS--outflow misalignment of
$0^\circ-25^\circ$, $R_{\rm{wind}} = 0.1-2.7\times 10^5~r_g =
0.4-9.6\times 10^{17}$~cm which suggests a physical scale of the order
of (or even slightly larger than) that of the BLR rather than of the
inner disc ($R_{\rm{BLR}} \sim 1-2\times 10^{16}$~cm according to
Kaspi et al. 2005 for the mean 2--10~keV luminosity of ESO~323--G77,
i.e. $\sim 5.8\times 10^{42}$~erg~s$^{-1}$).  It is therefore tempting
to associate the ionized gas with a warm/hot medium that may fill the
intra--cloud BLR, possibly enabling the cold, line--emitting clumps of
the BLR to be pressure--confined. 

\subsection{A relativistic, broad Fe K$\alpha$ line?}
\label{broadFe}

In their analysis of the {\it XMM--Newton} observation 1 on 2006/02,
Jim\'enez--Bail\'on et al. (2008a) report the detection of a
relativistically broadened Fe K$\alpha$ emission line. Their
best--fitting model comprises a broad $6.4$~keV line off the inner
accretion disc of a spinning Kerr black hole viewed at $\sim 26^\circ$
inclination. Adding a {\small{LAOR}} line model with rest--frame
energy fixed at $6.4$~keV to our best--fitting solution for the same
observation, the fitting statistic improves significantly with $\Delta
\chi^2 = 17$ for 4 additional free parameters, and most of the
residuals shown in the lower panel of Fig.~\ref{ratios} around 6--7~keV
are accounted for. We confirm their main findings, and we measure a
line equivalent width of $280\pm 180$~eV with respect to the intrinsic
nuclear continuum, a disc inclination of $26^\circ\pm 7^\circ$ and a
disc inner radius of $\leq 10~r_g$, for which any black hole spin is
allowed. We measure a rather typical radial emissivity index of
$2.5\pm 0.5$. We point out that the remaining parameters of the global
model are all consistent with the results of Table~\ref{tab2} within
the quoted uncertainties.

We then apply the same model to the multi--epoch data forcing, for
simplicity, all line parameters except the normalisation to be the
same in all data sets with sufficiently high quality, namely the {\it
  XMM--Newton} observations 1 and 10, and the {\it Chandra}
observations 5--8. We do not consider the good quality {\it Suzaku}
data here, because the Compton--thick absorption towards the X--ray
source prevents the detection of any contribution from the innermost
accretion flow. The fitting statistic before applying the broad line
model is $\chi^2 = 2410$ for 2435 dof, and the improvement obtained
with the addition of the broad line is $\Delta \chi^2 = 22$ for 6
additional free parameters. However, the line is not formally detected
during the 2010/04 {\it Chandra} observation 5--8, and is only
marginally detected during the 2013/01 {\it XMM--Newton} observation
10. All broad line parameters are consistent with those derived for
the {\it XMM--Newton} observation 1. The resulting line equivalent
widths are $280\pm 180$~eV, $\leq 108$~eV, and $60\pm 50$~eV for the
{\it XMM--Newton} observation 1, {\it Chandra} observation 5--8, and
{\it XMM--Newton} observation 10, respectively. Hence the data are
marginally consistent with a constant line equivalent width of $\sim
100$~eV, which would indicate that the line responds to the intrinsic
continuum variation. We must, however, point out that while a broad
feature that is consistent with a relativistic Fe line is indeed
detected in the {\it XMM--Newton} observations 1 and 10 and is allowed
in the {\it Chandra} observation 5--8, the complexity of our spectral
model prevents us to claim the secure detection of a broad
relativistic line in ESO~323--G77, as other interpretation of this
spectral feature may be possible (e.g. Turner \& Miller 2009).

\begin{figure*}
\begin{center}
\includegraphics[width=0.5\textwidth,height=0.7\textwidth,angle=-90]{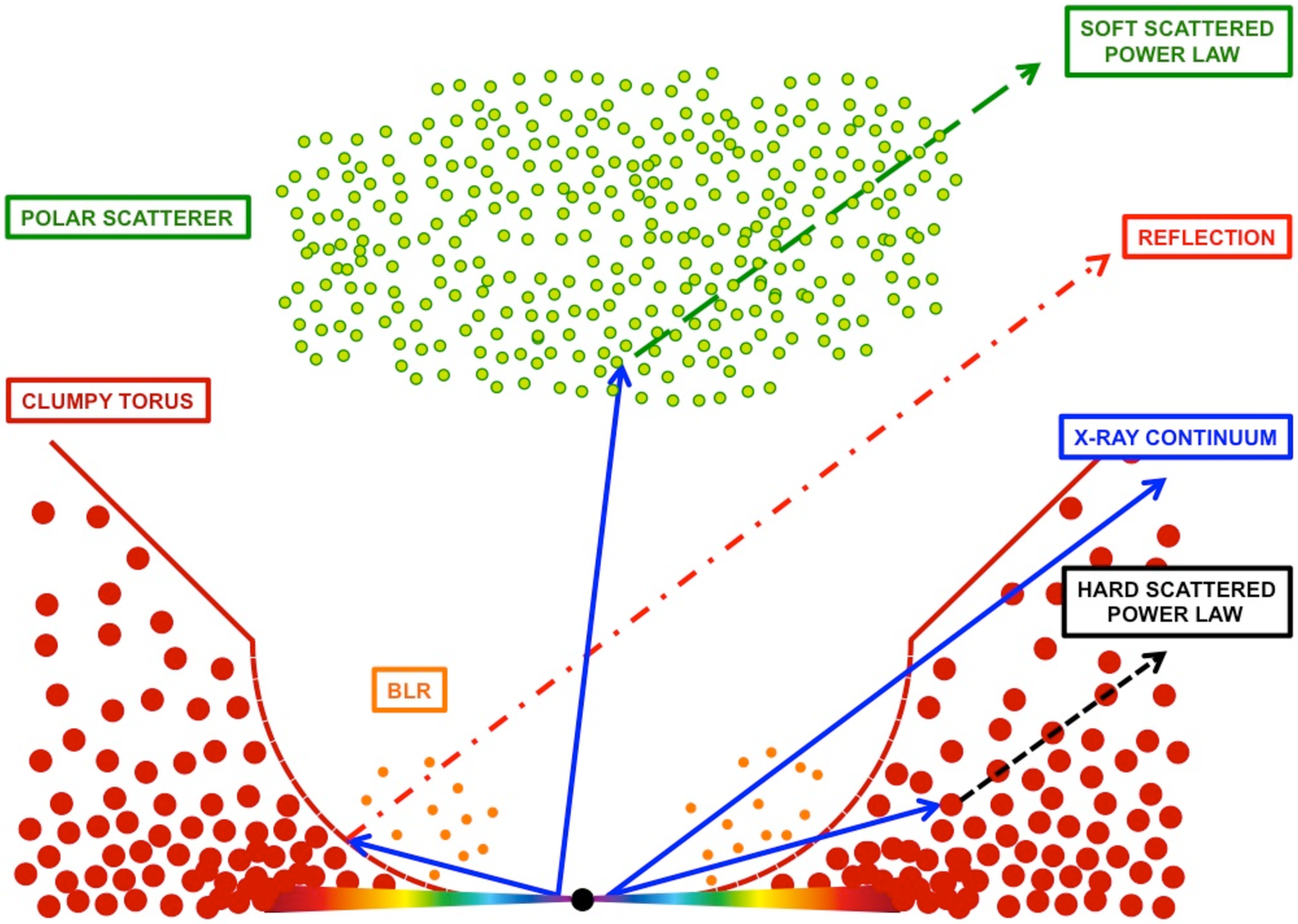}
\caption{A sketch of a possible geometry for the systems of absorbers,
  as suggested by the X--ray data. We assume an inclination angle of
  $\sim 45^\circ$, and our LOS is on the right--hand--side of the
  figure. The intrinsic nuclear continuum is shown with solid lines
  (blue in the on--line version) and goes through one cloud of the
  clumpy torus. This situation is therefore appropriate for
  observations 1, 3, 4, and 5-8. The remaining observations correspond
  instead to one of the BLR clouds crossing our LOS. The continuum
  illuminates a polar scattering region which gives rise to the soft
  scattered power law component which is likely associated with part
  of the soft emission lines spectrum (long--dashed line, green in the
  on--line version). For simplicity, we do not show star--forming
  regions that may be associated with the plasma emission model. The
  continuum also illuminates the far side of the clumpy torus which,
  at least at low latitudes, is likely optically--thick, producing the
  X--ray reflection spectrum (dot--dashed line, red in the on--line
  version).  Finally, the X--ray continuum intercepts clouds in the
  clumpy torus (and/or the BLR) out of our LOS. These clouds scatter
  part of the irradiating continuum into our LOS giving rise to a hard
  scattered power--law (short--dashed black line) that is absorbed by
  a different column density than the X--ray nuclear
  continuum, as it generally goes through a different system of
  clouds.}
\label{geometry}
\end{center}
\end{figure*}

\subsection{The scattered component}

As already discussed in Section~\ref{scattexpla} we interpret the
additional absorbed power law that we need to account for the spectral
shape of the {\it XMM--Newton} observation 10 as a scattering
component originating in a clumpy absorber. Our analysis of the X--ray
absorption variability in ESO~323--G77 confirms the presence of such
clumpy absorber and therefore supports our interpretation (see also
the Section~\ref{discuss}). Notice that a similar scattered component
was suggested also in NGC~7582 (Piconcelli et al. 2007) and NGC~3277
(Lamer, Uttley, McHardy 2003), two sources where X--ray absorption
variability is also present suggesting the presence of a clumpy
absorber (most likely the BLR).

As mentioned, the amount of optical polarisation as well as the
orientation of the position angle with respect to the [O~\textsc{iii}]
ionization cone, strongly suggests that we are seeing the nucleus with
$\sim 45^\circ$ inclination. Assuming that the half--opening angle of
the clumpy torus is similar, our LOS likely intercepts the torus
atmosphere. A sketch of the envisaged nuclear geometry
in ESO~323--G77 is shown in Fig.~\ref{geometry}. In such situation,
scattering can arise in the clumpy torus itself, from LOS with higher
inclination than ours, so that the scattered component emission
reasonably intercepts more clumps than the intrinsic nuclear continuum
(e.g. the situation shown in Fig.~\ref{geometry}). Indeed, we measure
a column density of $N_{\rm H}^{\rm scatt} = 7-8 \times
10^{22}$~cm$^{-2}$ towards the scattered component, and a (variable)
column density of $N_{\rm H}^{\rm nucl} = 2-6 \times
10^{22}$~cm$^{-2}$ towards the nuclear continuum, possibly indicating
that the scattered component goes through more absorbing clouds than
the nuclear emission into our LOS. Notice that scattering can also
arise from BLR clouds and not only from the clumpy torus
itself. 

The scattered component we propose in in fact detectable only in cases
where the intrinsic nuclear continuum is more absorbed than the
scattered component itself. Hence, we do not expect to detect this
component in typical unabsorbed Seyfert~1 galaxy, although it may be
ubiquitous in the X--ray spectra of AGN. On the other hand, we suggest
that the scattered component should be detected in all Seyfert~1
galaxies whose nuclear continuum is transiently obscured by clouds
with column density in excess of ${\rm{few}}\times
10^{22}$~cm$^{-2}$. As for genuine Compton--thin and/or Compton--thick
Seyfert~2 galaxies with highly inclined LOS, the scattered component
should still be present, but significantly more absorbed than in
ESO~323--G18 which may yield to its detection in some cases, but
probably not in those where the nucleus is seen edge--on (or nearly
edge--on) where the difference in column density between the nuclear
continuum and the scattered component may be too small.

In principle, the scattered component is likely to contribute some
line emission in the Fe K band which may be variable if scattering
occurs off clouds that are close to our LOS. However, if the scattered
flux is only $\sim 15$~\% of the intrinsic (see Table~\ref{tab2}), the
Fe line flux likely comprises $\leqsim 15$~\% contribution from the
scattered component. Our data are consistent with a constant Fe line
flux, but we cannot exclude the presence of a small variable component
of the Fe emission line at such a fine level (as an example, the Fe
line flux is only constrained at the $\sim 25$~\% level in the {\it
  XMM--Newton} observation 1.)

As for the reflection component, the Fe line width of $0.06$~keV
(corresponding to a FWHM$\leq 6600$~km~s$^{-1}$) is consistent with
any possible physical scale from the inner BLR outwards. In
Fig.~\ref{geometry} we show only one possible origin for such
component, i.e. the far side of the clumpy torus at low elevation,
where the dusty gas is likely optically--thick. However, our analysis
cannot rule out that reflection arises (or also arises) in material
that is either closer (e.g. the BLR or the near side of the torus) or
farther away (see e.g. Iwasawa et al 2003; Guainazzi et al. 2012) than
the far side of the clumpy torus itself.

\begin{figure*}
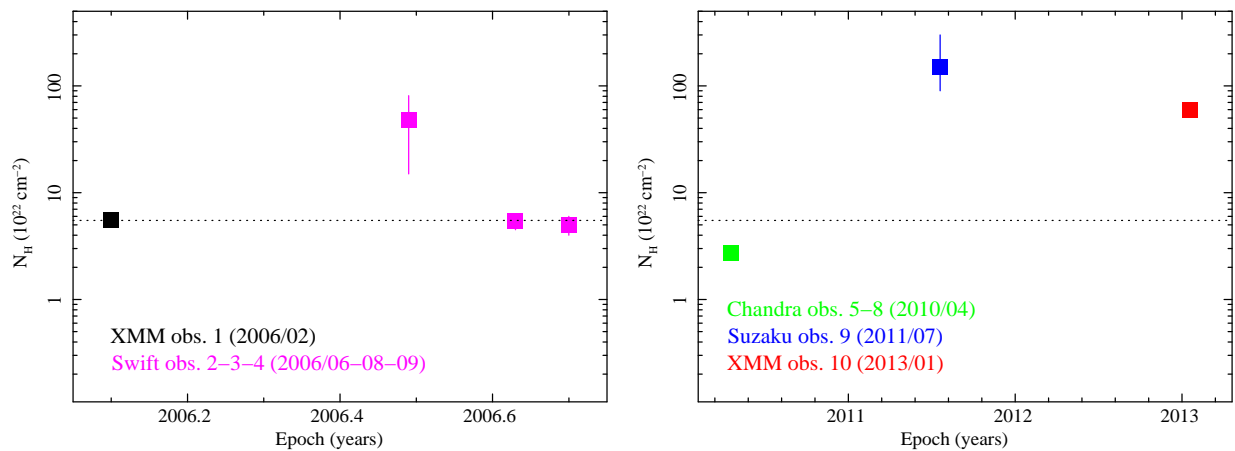

\begin{center}
{
\mbox{\includegraphics[width=0.33\textwidth,height=0.45\textwidth,angle=-90]{nh1.ps}}
{\hspace{0.2cm}}
\mbox{\includegraphics[width=0.33\textwidth,height=0.45\textwidth,angle=-90]{nh2.ps}}
}
\caption{Column density of the variable neutral absorber as a function
  of observation epoch. The dotted line is the best--fitting column density to
  observations 1, ,3, and 4 that appear to be absorbed by the same
  column of neutral gas.}
\label{nh}
\end{center}
\end{figure*}

\section{The variable absorber}
\label{discuss}

Fig.~\ref{nh} shows the column density towards the intrinsic nuclear
continuum as a function of time, summarising the most important result
of our multi--epoch spectral analysis, namely the remarkable X--ray
absorption variability towards the nuclear continuum in
ESO~323--G77. The 2006 observations (1 to 4 in Table~\ref{tab1}) are
shown in the left panel, and the remaining 2010--2013 ones in the right
panel. The less X--ray absorbed state is observed during the {\it
  Chandra} monitoring from 2010--04--14 to 24 where the source appears
to be seen through a column of $\sim 2.7\times 10^{22}$~cm$^{-2}$. On
the other hand, the {\it Suzaku} observation of 2011--07--20 is
reflection--dominated and strongly suggests a Compton--thick state
with a neutral absorber column density of $\sim 1.5\times
10^{24}$~cm$^{-2}$. The other observations reveal
intermediate--absorption states with neutral column densities ranging
from $\sim 5\times 10^{22}$~cm$^{-2}$ to $\sim 6\times
10^{23}$~cm$^{-2}$.

The observed X--ray absorption variability allows us to study the
properties of the absorbing systems in great detail. Let us assume the
simplest possible geometry, namely an X--ray source with linear size
$D_{\rm s}$ and an absorber (cloud) of uniform column density and size
$D_{\rm c}$ (e.g. a sphere of diameter $D_{\rm c}$). As our
multi--epoch spectral analysis never reveals a partial covering
structure, $D_{\rm c} \geq D_{\rm s}$. We also assume that the
absorbing cloud moves with transverse velocity $v_{\rm c}$ with
respect to our LOS. The most important timescale that can be derived
from our analysis is $\Delta T_{\rm{const}}$, i.e. the timescale over
which the X--ray source is covered by the same absorbing cloud.
$\Delta T_{\rm{const}}$ can be related to $D_{\rm c}$, $D_{\rm s}$,
and $v_{\rm c}$ via $D_{\rm c} = \Delta T_{\rm{const}} v_{\rm c} + D_{\rm s}$, so that 
\begin{equation}
\Delta T_{\rm{const}}^{\rm{min}} v_{\rm c}^{\rm{min}} + D_{\rm s} \leq D_{\rm c} \leq \Delta T_{\rm{const}}^{\rm{max}} v_{\rm c}^{\rm{max}} + D_{\rm s} \, . 
\label{const} 
\end{equation}
As for the X--ray source size, we impose $D_{\rm{s}} \simeq 20~r_g$,
in agreement with previous X--ray occultation results (e.g. NGC~1365,
see Risaliti et al. 2007) and with micro--lensing results (e.g. Dai et
al. 2010; Mosquera et al. 2013). For a black hole mass of $M_{\rm{BH}}
\simeq 2.5\times 10^7~M_\odot$ (e.g. Wang \& Zhang 2007), one has
$D_{\rm{s}} \simeq 7.4\times 10^{13}$~cm.
 
Hereafter, we critically consider three possible origins for the
variable absorber, namely (i) the classical torus of Unified models;
(ii) the BLR; (iii) a disc--wind launched off the
inner accretion disc. The three possible origins are spatially
located at different scales and, under the reasonable assumption that
the cloud transverse motion is dominated by gravity (except for the
wind, where the wind outflow velocity plays a role), they are
associated with different typical values for the cloud velocity.

{\bf The torus:} the dusty torus inner boundary is identified with the
dust sublimation radius $R_{\rm{dust}} \simeq 0.4 L_{45}^{0.5}$~pc
(Elitzur \& Shlosman 2006), where $L_{45}$ is the bolometric
luminosity in units of $10^{45}$~erg~s$^{-1}$. The averaged intrinsic
luminosity of ESO~323--G77 is $\sim 5.8\times 10^{42}$~erg~s$^{-1}$ in
the 2--10~keV band. Assuming a 2--10~keV bolometric correction of 20,
as appropriate for AGN with Eddington ratio $\leq 0.1$ (e.g. Vasudevan
\& Fabian 2007), gives $L_{\rm{Bol}} \sim 1.2\times
10^{44}$~erg~s$^{-1}$ and $R_{\rm{dust}} \simeq 0.14$~pc. As for the
torus outer edge, current IR observations indicate that the outer edge
of the torus is $R_{\rm{out}} \simeq 5-30~R_{\rm{dust}}$
(e.g. Nenkova, Ivezi{\'c} \& Elitzur 2002; Poncelet et al. 2006;
Tristram et al. 2007; Nenkova et al. 2008b; Ramos Almeida 2011;
Alonso--Herrero 2011). As we are interested in a conservative upper
limit, we assume that $R_{\rm{out}} = 5$~pc (corresponding to $\sim
35\times R_{\rm{dust}}$). Assuming Keplerian motion, one has $v_{\rm
  c}^{\rm{TOR}} = 150-900$~km~s$^{-1}$.

{\bf The BLR:} a natural upper limit on the BLR location is the
dust--sublimation radius itself. In fact, the BLR and torus are most
likely part of the same obscuring structure which simply changes from
gas-- to dust--dominated at $R_{\rm{dust}}$ (e.g. Elitzur \& Shlosman
2006), so that $v_{\rm c}^{\rm{BLR, min}} = 900$~km~s$^{-1}$. As for
the BLR inner maximum velocity, let us consider the observed broad
H$\beta$ line FWHM$\simeq 2500$~km~s$^{-1}$ (Winkler 1992), which is
related to the actual cloud velocity by the BLR geometry and/or by our
particular viewing angle. Assuming a spherical BLR, this translated
into a cloud typical velocity $\simeq 2200$~km~s$^{-1}$. For a
disc--like, flattened BLR, the relationship between FWHM and actual
velocity depends on the viewing angle, the velocity being higher for
lower inclination angles of our LOS.  As mentioned, there are
important indications from spectro--polarimetry that our viewing angle
is close to $45^\circ$ in ESO~323--G77. In order to derive the most
conservative cloud velocity upper limit, we assume here a lower
inclination of $25^\circ$, which corresponds to $v_{\rm c}^{\rm{BLR,
    max}} = 3000$~km~s$^{-1}$. Summarising, we have $v_{\rm
  c}^{\rm{BLR}} = 900-3000$~km~s$^{-1}$.

{\bf The inner disc--wind:} in recent years, the presence of a
disc--wind launched off the inner accretion flow has been often
invoked to account for spectral and variability properties of AGN in
the X--ray band (e.g. Miller et al. 2010; Tombesi et al. 2010, 2012;
Tatum et al. 2012). Theoretical modelling suggests that disc--winds
are launched off radii of the order of $100-700~r_g$ on the accretion
disc (Risaliti \& Elvis 2010; Nomura 2013). The wind clumps transverse
motion is likely dominated by gravity, although depending on the
outflow direction with respect to the LOS, the outflow velocity may
also play a significant role. Assuming that the latter equals the
escape velocity at the launching radius, we can use the escape
velocity at $100~r_g$ as an upper limit on the cloud velocity, and the
Keplerian velocity at $700~r_g$ as a lower limit to infer that $v_{\rm
  c}^{\rm{WIND}} = 11000-42000$~km~s$^{-1}$.

\subsection{The ${\rm{few}}\times 10^{22}$~cm$^{-2}$ absorber during observations 1, 3, 4, and 5-8}

Let us first focus our attention on the observations absorbed by
column densities of ${\rm{few}}\times 10^{22}$~cm$^{-2}$, namely
observations 1, 3, 4, and 5--8. As shown in Fig.~\ref{nh}, the 2006
observations 1, 3, and appear to be absorbed by the same column
density ($\sim 5.5\times 10^{22}$~cm$^{-2}$), while a sudden increase
of the column density [$\Delta N_{\rm H} = (4.5\pm 1.8)\times
  10^{23}$~cm$^{-2}$] is observed during observation 2.  It is
tempting to regard the overall 2006 column density variability as due
to a constant--column--density absorber with $N_{\rm H}\sim 5.5\times
10^{22}$~cm$^{-2}$ covering the X-ray source during the 7 months
spanned by observations 1 to 4 (2006/02 to 2006/09) plus an additional
obscuration event taking place on 2006/06 during observation 2. On the
other hand, the 2010/04 observation 5--8 (3.6~yr after observation 4)
appears to be absorbed by a slightly lower column density of $N_{\rm
  H}\sim 2.7\times 10^{22}$~cm$^{-2}$.

The lack of absorption variability in 2006 (besides the event on
2006/06, observation 2) suggests that $\Delta T_{\rm{const}} \geq
7$~months. On the other hand, no information of the upper limit on
$\Delta T_{\rm{const}}$ is available, due to the lack of data before
2006. Using Eq.~\ref{const} with the various $v_{\rm c}$ that we have
derived in the previous Section and with $D_{\rm s} = 7.4\times
10^{13}$~cm, one has $D_{\rm c}^{\rm{TOR}} \geq 3.5\times 10^{14}$~cm,
$D_{\rm c}^{\rm{BLR}} \geq 1.7\times 10^{15}$~cm, and $D_{\rm
  c}^{\rm{WIND}} \geq 2.0\times 10^{16}$~cm. By combining the cloud
size(s) with the maximum common column density during the 7 months
corresponding to observations 1, 3, and 4 (namely $6\times
10^{22}$~cm$^{-2}$), we obtain upper limits on the cloud number
density for the three cases, i.e. $n_{\rm c}^{\rm{TOR}} \leq 1.7\times
10^8$~cm$^{-3}$, $n_{\rm c}^{\rm{BLR}}\leq 3.5\times 10^7$~cm$^{-3}$,
and $n_{\rm c}^{\rm{WIND}} \leq 3.0\times 10^6$~cm$^{-3}$.

Only $n_{\rm c}^{\rm{TOR}}$ is consistent with the density that is
expected in the dusty torus (e.g. Elitzur \& Shlosman 2006), while the
upper limits on $n_{\rm c}^{\rm{BLR}}$ and $n_{\rm c}^{\rm{WIND}}$ are
at least two orders of magnitude lower than the number density of the
BLR and/or disc--wind. We conclude that the ${\rm{few}}\times
10^{22}$~cm$^{-2}$ absorber detected during observations 1, 3, 4, and
5--8 is associated with the torus at distances of $R_{\rm
  c}^{\rm{TOR}} = 0.14-5$~pc from the central black hole. On the other
hand absorber is variable on long timescales ($\geq 7$~months), as
shown by the lower column density during the 2010/04 {\it Chandra}
observation 5--8. Such variability implies that the torus is clumpy
rather than homogeneous, an important X--ray--only result which
confirms previous indications for the clumpiness of the parsec--scale
obscuring torus obtained in the IR (e.g. Ramos Almeida et al. 2011).

The column density of the individual clouds of the clumpy torus (say
the observed mean of $\sim 4\times 10^{22}$~cm$^{-2}$) can also be
used to estimate the critical LOS with inclination $i_{\rm{crit}}$
that separates dust--free and dust--rich regions. Assuming the
formalism of Liu \& Zhang (2011), the critical LOS depends on the
Eddington ratio ($\sim 0.036$ for ESO~323--G77) and on the radiation
pressure boost factor for dusty gas $A$ (i.e. the ratio of the
absorption cross section for dusty gas to that for electrons
alone). For the given Eddington ratio, and for clouds with $N_{\rm H}
\sim 4\times 10^{22}$~cm$^{-2}$, $A\sim 20$ (Fabian, Vasudevan \&
Gandhi 2008) so that $i_{\rm{crit}} \sim 47^\circ$ (see Eq.~3 of Liu
\& Zhang 2011). We then conclude that the half--opening angle of the
clumpy, dusty torus in ESO~323--G77 is of the order of $i_{\rm{crit}}
\sim 47^\circ$ in line with our expectation that our LOS (likely at
$i\sim 45^\circ$) grazes the torus and intercept its
atmosphere. Higher (lower) Eddington ratio AGN with torii having
clumps of similar column density have larger (smaller) half--opening
angles.

\subsection{The $10^{23}-10^{24}$~cm$^{-2}$ absorber during observations 2, 9, and 10}

We now focus on the more absorbed states observed on 2006/06
(observation 2), 2011/07 (observation 9), and 2013/01 (observation 10)
where column densities of $5.0\pm 1.8 \times 10^{23}$~cm$^{-2}$,
$1.5^{+1.5}_{-0.6} \times 10^{24}$~cm$^{-2}$, and $6.0 \pm 0.7\times
10^{23}$~cm$^{-2}$ respectively are measured, see Fig.~\ref{nh}. As
discussed in Section~\ref{varsec} (see Fig.~\ref{shortvar}), no
absorption variability is detected during the {\it XMM--Newton}
observation 10, so that $\Delta T_{\rm{const}} \geq 1.2\times 10^5$~s. On the
other hand, the sudden increase in column density during observation 2
can also be associated with a $10^{23}-10^{24}$~cm$^{-2}$ absorber
which is superimposed to the clumpy torus absorber discussed in the
previous section (see Fig.~\ref{nh}). From this event, one has that
$\Delta T_{\rm{const}}$ cannot be larger than the time--interval
between observations 1 and 3, i.e. $\Delta T_{\rm{const}} \leq
6.3$~months.

Using again Eq.~\ref{const}, we then have that $D_{\rm c}^{\rm{TOR}} =
0.8-14\times 10^{14}$~cm, $D_{\rm c}^{\rm{BLR}} = 0.9-50\times
10^{14}$~cm, and $D_{\rm c}^{\rm{WIND}} = 0.2-70\times 10^{15}$~cm
which, when combined with the observed common range of column
densities in observations 2 and 10 ($5.3-6.7\times 10^{23}$~cm$^{-2}$)
imply that $n_{\rm c}^{\rm{TOR}} = 0.4-8\times 10^9$~cm$^{-3}$,
$n_{\rm c}^{\rm{BLR}}= 0.1-8\times 10^9$~cm$^{-3}$, and $n_{\rm
  c}^{\rm{WIND}} = 0.008-3\times 10^9$~cm$^{-3}$.

The derived range of $n_{\rm c}^{\rm{TOR}}$ is inconsistent with the
upper limit derived in the previous section ($n_{\rm c}^{\rm{TOR}}
\leq 1.7\times 10^8$~cm$^{-3}$) so that we can exclude that the clumpy
torus is responsible for the absorption events with column density of
$10^{23}-10^{24}$~cm$^{-2}$, unless a very wide range of cloud
properties is assumed for the same LOS. On the other hand, $n_{\rm
  c}^{\rm{BLR}}$ is consistent with the number density needed to
produce the optical and UV broad emission lines in the BLR, so that an
identification of the absorber with a BLR cloud is totally
plausible. 

As for the wind, the upper limit on $n_{\rm c}^{\rm{WIND}}$ may be
marginally consistent with the expected wind density. However, some
further constraints can be derived from the fact that our spectral
analysis reveals absorption by neutral, rather than ionized gas. In
fact, if we replace the neutral absorber of our spectral analysis by a
ionized one, we have that $\log\xi \leq 1.3$. If we consider the
densest possible wind--cloud ($n_{\rm c}^{\rm{WIND}} = 3\times
10^{9}$~cm$^{-3}$) together with the upper limit on the ionization
parameter, then the wind must be located at $R_{\rm c}^{\rm{WIND}}\geq
2.7\times 10^{16}$~cm or $\geq 7.3\times 10^{13}~r_g$. Hence, we can
exclude that we are seeing absorption by an inner disc--wind at the
launching radii. On the other hand, the wind may participate to the
observed absorption variability once it reaches much larger distances,
that are typical of the BLR. In this framework, however, the wind and
BLR are indistinguishable.

We conclude that the variable absorber with column density of
$10^{23}-10^{24}$~cm$^{-2}$ observed during observations 2, 9 and 10
is consistent with a BLR origin. The properties of the variable
absorber(s) that we estimate from our analysis are reported in
Table~\ref{tab3}. We remind here that the only important assumptions
we have made are (i) that the X--ray source has size of $D_{\rm s}\ =
20~r_g$, and (ii) that the individual clouds have velocities
appropriate to be associated with the torus, the BLR, or a disc--wind
launched off the innermost few hundreds $r_g$.

\begin{table}
\caption{The derived properties of the absorbing clouds in the
  different observations, together with their subsequent
  identification.}
\label{tab3}      
\begin{center}
\begin{tabular}{l c c}
\hline\hline                 
\multicolumn{1}{l}{Cloud properties} & Obs. 1, 3, 4, 5--8 & Obs. 2, 9, 10 \\ \\
$N_{\rm H}$~[cm$^{-2}$] & $2-6\times 10^{22}$& $0.3-3\times 10^{24}$\\
$D_{\rm c}$~~[cm]& $\geq 3.5\times 10^{14}$& $0.9-50\times 10^{14}$\\
$n_{\rm c}$~~~[cm$^{-3}$]&$\leq 1.7 \times 10^{8}$& $0.1-8\times 10^{9}$ \\
$v_{\rm c}$~~~[km~s$^{-1}$] & $150-900$ &$900-3000$\\
\hline
Origin & Clumpy TORUS & BLR  \\
\hline\hline                        
\end{tabular}
\\
\end{center}
\end{table}

\subsection{Optical and UV simultaneous data from the {\it XMM--Newton} OM}

According to our interpretation of the two {\it XMM--Newton}
observations, the 2006/02 observation 1 is only absorbed by the clumpy
torus, while the 2013/01 observation 10 is absorbed by (or
additionally absorbed by) the BLR. In general, BLR clouds are not
expected to induce absorption variability in the UV as (i) the BLR are
most likely dust--free, and (ii) even if they were dusty (at least
partially, see e.g. Czerny \& Hryniewicz 2011) their small size with
respect to the UV--emitting region implies that they cannot imprint
strong UV absorption variability, as they always cover globally the
same fraction of the UV--emitting region. Hence, if observation 10 is
indeed absorbed by the BLR cloud, no strong UV variability is expected
between the two observations. Conversely, if the large column density
during observation 10 occurs in the clumpy, dusty torus one may expect
a lower UV flux during observation 10 than during observation
1. Although we have already excluded that this is the case, our result
can be checked by looking for UV variability between observations 1
and 10.

We derive the UV flux densities in the {\it XMM--Newton} OM filters W2
($2120$\AA), M2 ($2310$\AA), and W1 ($2910$\AA) for both observations
(filters V and U where used in addition during observation 1 and 10
respectively, so that they cannot be compared). Our results are
reported in Table~\ref{tab4}. As clear from the table, the UV fluxes
during the more absorbed observation 10 are slightly higher than those
during the less absorbed observation 1. Intrinsic flux variability
cannot play a significant role, has the X--ray luminosity is in fact
lower during observation 10 (see Table~\ref{tab2}). This rules out
that the $\sim 6\times 10^{23}$~cm$^{-2}$ column density detected
during observation 10 is due to a dusty extended medium (e.g. the
clumpy torus) which affects the UV--emitting region of the disc, and
strongly supports our interpretation in terms of BLR absorption.

\begin{table}
\caption{Flux densities in the UVW2, UVM2, and UVW1 OM filters during
  the {\it XMM--Newton} observations 1 and 10 are given in units of
  $10^{-15}$~erg~cm$^{-2}$~s$^{-1}$~$\AA^{-1}$.}
\label{tab4}      
\begin{center}
\begin{tabular}{l c c}
\hline\hline                 
OM filter & $f^{\rm{UV}}_{\rm{obs.1}}$& $f^{\rm{UV}}_{\rm{obs.10}}$ \\ \\
UVW2 & $1.5\pm 0.1$ & $1.8\pm 0.1$ \\
UVM2 & $1.6\pm 0.1$ & $2.3\pm 0.1$ \\
UVW1 & $4.4\pm 0.2$ & $5.2\pm 0.3$ \\
\hline\hline                        
\end{tabular}
\\
\end{center}
\end{table}

\section{Clumpy absorbers in AGN: comparison with previous studies}

Clumpy X--ray absorbers have been previously detected in other AGN via
X--ray absorption variability studies. In general, short--timescale
absorption variability events have been associated with compact
absorbers at the BLR scale, while (rarer) long--timescale events have
been attributed to more extended absorbers such as the clumpy
torus. Here we briefly compare our results on ESO~323--G77 with some
of previous ones.

{\bf Clumpy torus absorption:} Rivers, Markowitz \& Rothschild (2011)
report the detection of an X--ray eclipse in Cen~A whose properties
are consistent with the transit of one cloud with column density $\sim
4\times 10^{22}$~cm$^{-2}$ into the LOS over the course of $\sim
5.7$~months. They were able to estimate a cloud linear size of
$1.4-2.4\times 10^{15}$~cm and a central number density of
$1.8-3.0\times 10^7$~cm$^{-3}$, in good agreement with previous
results on the same source (Rothschild et al. 2011). The properties
of the absorber imply an absorption event associated with the clumpy
torus and are consistent with those we derive in ESO~323--G77. This is
true for the observed column density of the individual clouds, as well
as for their size and number density (see Table~\ref{tab3}). This strongly
favours models in which the torus is made of individual clumps of
column density of ${\rm{few}}\times 10^{22}$~cm$^{-2}$ and number density
of the order of ${\rm{few}}\times 10^{7}$~cm$^{-3}$, as suggested by
theoretical models (e.g. Nenkova et al. 2008a, 2008b).

{\bf BLR:} X--ray absorption variability on much shorter timescales
has been also reported in a number of sources. The most remarkable
case is that of the Seyfert 1.8 galaxy NGC~1365 which shows common,
multiple, and very rapid absorption variability events on timescales
as short as a few hours (Risaliti et al. 2005; 2007; 2009). Adequate
monitoring of these events allowed Risaliti et al. (2007) to constrain
the X--ray continuum source size to $20~r_g$ at most (Risaliti et
al. 2007; 2009), and to infer that absorption occurs in clouds with
density of the order of $10^{10}$~cm$^{-3}$, size of few $r_g$, and
velocity typical of the BLR. Further studies on the same source
allowed to study the structure of the absorber in great detail, and
Maiolino et al. (2010) conclude that the clouds have a strongly
elongated and cometary shape, with a dense head, and an expanding,
dissolving tail. The cometary tail inferred in NGC~1365 may well be a
general property of the BLR clouds. As for ESO~323--G77, the cometary
tail could be related to the highly ionized absorbers we detect,
although the lower quality of our data with respect to those used by
Maiolino et al. (2010) prevents us from performing further tests of
this hypothesis. Other examples of X--ray absorption by clouds that
are consistent with a BLR origin include Mrk~766 (Risaliti et
al. 2011), where clouds with column density of ${\rm{few}}\times
10^{23}$~cm$^{-2}$, density of $10^{10}-10^{11}$~cm$^{-3}$, and linear
size of $5-50~r_g$ are inferred, and SWIFT~J2127.4+5654 (Sanfrutos et
al. 2013) where the absorbing cloud has a lower column density of
${\rm{few}}\times 10^{22}$~cm$^{-2}$, a density of $\geq 1.5\times
10^9$~cm$^{-3}$ and a linear size of few $r_g$. In the latter case,
the X--ray coverage of a full partial X--ray eclipse allowed Sanfrutos
et al. (2013) to also estimate the X--ray continuum source linear size
to be $\leq 10.5~r_g$. The BLR cloud(s) properties we estimate in
ESO~323--G77 are consistent with the above results (see
Table~\ref{tab3}), although the slightly lower number density we
obtain may indicate that absorption in ESO~323--G77 occurs in the
outer BLR, closer to the dust sublimation radius than in NGC~1365 or
Mrk~766. In summary, X--ray absorption variability on
short--timescales points towards the presence of a compact (few $r_g$)
X--ray continuum source that can be transiently obscured by clouds
associated with the BLR.

{\bf Variable absorption in polar--scattered AGN:} Finally, it is
worth mentioning that X--ray absorption variability on a variety of
timescales has been reported in a number of other polar--scattered AGN
such as NGC~3227 (Lamer et al. 2003), Mrk~1239 (Grupe \& Mathur 2004),
Mrk~704 (Matt et al. 2011), Mrk~766 (Risaliti et al. 2011), and
Mrk~231 (Piconcelli et al. 2013). Preliminary analysis reveals that
absorption variability is also present in the polar--scattered
Seyfert~1 galaxies Fairall~51 and (although somewhat marginally)
NGC~4593 (Miniutti et al. in preparation; see also
Jim{\'e}nez-Bail{\'o}n et al. 2008b). The large number of
polar--scattered Seyfert~1 galaxies showing X--ray absorption
variability (7 out of the 12 polar--scattered Seyfert~1 galaxies
defined in the sample by Smith et al. 2004) strongly suggests that
intermediate--inclination LOS of $40^\circ-50^\circ$ are highly likely
to give rise to X--ray absorption variability either due to clouds in
clumpy torus or to BLR clouds, which may be out of the LOS for
significantly less inclined LOS.

\section{Summary and conclusions}

We report results from 10 X--ray observations of the polar--scattered
Seyfert~1 galaxy ESO~323--G77 performed in the 2006--2013 time
frame. The source exhibits remarkable X--ray spectral variability that
is unambiguously associated with the variation of the column density
of a neutral absorber towards the nuclear continuum. Our multi--epoch
spectral analysis of ESO~323--G77 allows us to identify X--ray
absorption by a clumpy torus, by BLR clouds, and by a warm/hot
outflowing medium. The properties of the absorbers can be summarised
as follows:

\begin{enumerate}
\item {\bf Clumpy torus:} the clumpy torus is responsible for X--ray
  absorption with column density of the order of $2-6\times
  10^{22}$~cm$^{-2}$. The torus clumpiness is demonstrated by the
  X--ray absorption variability on timescales shorter than $3.6$~yr
  and longer than a few months. Our data enable us to infer that the
  typical size of the clumpy torus clouds is $D_{\rm c}^{\rm{TOR}}
  \geq 3.5\times 10^{14}$~cm with a typical number density $n_{\rm
    c}^{\rm{TOR}} \leq 1.7\times 10^8$~cm$^{-3}$. By combining the
  observed clumps column density with the radiation pressure boost
  factor for dusty gas, we estimate a torus half--opening angle of
  $\sim 47^\circ$. Although we are likely probing only the torus
  atmosphere with our X--ray observations, it is worth mentioning that
  if the cloud properties that we infer are representative of the
  typical clumpy torus cloud, a relatively large number $N_{\rm c}
  \geq 15$ of clouds is required to be present in the equatorial plane
  to account for Compton--thick Seyfert~2 galaxies seen at nearly
  edge--on inclinations. 
\item {\bf BLR clouds:} BLR clouds are instead responsible for X--ray
  absorption with column density of the order of $0.3-3.0\times
  10^{24}$~cm$^{-2}$. The X--ray variability can be used to infer a
  typical BLR cloud size $D_{\rm c}^{\rm{BLR}} = 0.9-50 \times
  10^{14}$~cm with a typical number density $n_{\rm c}^{\rm{BLR}} =
  0.1-8\times 10^9$~cm$^{-3}$, in line with the density that is needed
  to efficiently produce the optical/UV broad emission lines. The
  derived number density is slightly lower than that inferred from
  occultation events in e.g. NGC~1365 and Mrk~766 (e.g. Risaliti et
  al. 2009; 2011) and may imply that absorption in ESO~323--G77 occurs
  in the outer BLR, closer to the dust sublimation radius than in
  NGC~1365 and Mrk~766.
\item {\bf Highly--ionized outflow:} we confirm the previous detection
  of a system of outflowing highly--ionized absorbers. The outflow
  velocity is of the order of $1000-4000$~km~s$^{-1}$, of the same
  order of the optical broad lines FWHM ($2500$~km~s$^{-1}$). If the
  outflow velocity is comparable to the Keplerian velocity at the
  launching radius (notice that the escape velocity is just a factor
  $\sqrt{2}$ of larger than the Keplerian), the wind is then
  consistent with being roughly co--spatial with the BLR. Therefore,
  it is tempting to identify the outflowing gas with the inter--cloud
  medium that provides the pressure confinement for the colder BLR
  clouds. The lack of significant variability of the warm absorbers
  may indicate a clumpy outflow whose warm/hot phase is nearly
  homogeneous, and whose clumps may be identified with the
  line--emitting cold, dense BLR clouds. The warm/hot phase may also
  be connected with the mass loss of the cold clumps, as suggested by
  the BLR cometary shape suggested by Maiolino et al. (2010) in the
  case of NGC~1365.
\end{enumerate}

In summary, multi--epoch X--ray observations of ESO~323--G77 enable us
to directly map the clumpy torus and BLR clouds in X--ray
absorption. We use X--ray data to derive with good precision the
main properties of the absorbers (such as the size and density of the
individual clumps). A more continuous X--ray monitoring of ESO~323--G77 and
similar sources will allow us to significantly refine the X--ray
derived clumpy torus and BLR properties in the future. We also suggest
an identification of the highly--ionized outflowing warm absorber in
ESO~323--G77 with the smooth, likely homogeneous inter--cloud BLR
medium which may provide the pressure confinement for the cold BLR
clouds. Within this scenario, the BLR clouds would simply be the cold,
dense clumps of the smoother warm/hot outflow.

\section*{Acknowledgements}

GM thanks Pilar Esquej and Almudena Alonso--Herrero for useful
discussions. This work is based on observations obtained with XMM-Newton, an ESA
science mission with instruments and contributions directly funded by
ESA Member States and NASA. This research has also made use of data
and/or software provided by the High Energy Astrophysics Science
Archive Research Center (HEASARC), which is a service of the
Astrophysics Science Division at NASA/GSFC and the High Energy
Astrophysics Division of the Smithsonian Astrophysical Observatory. We
made use of data obtained from the Chandra Data Archive and the
Chandra Source Catalogue, and software provided by the Chandra X-ray
Center (CXC) and of data obtained from the Swift observatory. We also
made use of data from the {\it Suzaku} observatory, a collaborative
mission between the space agencies of Japan (JAXA) and the USA
(NASA). BAG, GM, and MS thank the Spanish MINECO for support through
the Spanish Plan Nacional de Astronom\'{\i}a y Astrof\'{\i}sica under
grant AYA2010-21490-C02-02.

\end{document}